\documentclass{aa}
\usepackage{natbib}\usepackage{txfonts}
\usepackage{graphicx}

\newcommand{\vi}{\vec{v}}

\newcommand{\lp}{ \;\left(}
\newcommand{\rp}{ \right)}
\newcommand{\nab}{ \vec{\nabla} }
\newcommand{\lapl}{ \Delta }

\newcommand{\noi}{ \noindent }

\newcommand{\dnz}[1]{\frac{d  #1}{d\zeta}}
\newcommand{\ddnz}[1]{\frac{d^2  #1}{d\zeta^2}}

\newcommand{\beq}{\begin{equation}}
\newcommand{\eeq}{\end{equation}}

\newcommand{\ltex}[1]{\quad \hbox{#1} \quad}
\newcommand{\llp}{ \ell(\ell+1)}
\newcommand{\lc}{ \left[}
\newcommand{\rc}{ \right]}
\newcommand{\greq}{\begin{equation} \begin{array}{l}}
\newcommand{\egreq}{\end{array} \end{equation}}

\newcommand{\YL}{ Y^m_\ell }
\newcommand{\YLM}{ Y^{m'}_\ell } 
\newcommand{\YTL}{\hat{Y}^m_\ell} 
\newcommand{\YTLP}{\hat{Y}^m_{\ell '}}
\newcommand{\DYTLP}{\frac{\partial \hat{Y}^m_{\ell '}}{\partial \theta}}
\newcommand{\lpp}{ \ell \ell '}

\newcommand{\eeqn}[1]{\label{#1}\end{equation}}

\newcommand{\eq}[1]{(\ref{#1})}     
\newcommand{\gi}{{\mbox{I}}^m_{\lpp}}
\newcommand{\gj}{{\mbox{J}}^m_{\lpp}} 
\newcommand{\beqa}{\begin{eqnarray*}}
\newcommand{\eeqa}{\end{eqnarray*}}
\newcommand{\beqan}{\begin{eqnarray}}
\newcommand{\eeqan}[1]{\label{#1}\end{eqnarray}}

\newcommand{\kms}{$\mathrm {km s}^{-1}$}
                                   
\begin{document}

\title{Acoustic oscillations in rapidly rotating polytropic stars}
\subtitle{I. Effects of the centrifugal distortion}
\author{F. Ligni\`eres \and M. Rieutord \and D. Reese}
\institute{
Laboratoire d'Astrophysique de Toulouse et Tarbes, UMR 5572, Observatoire Midi-Pyr\'en\'ees, Universit\'e Paul Sabatier - Toulouse 3,
14 avenue \'E. Belin, 31400 Toulouse,
France \\
\email{francois.lignieres@ast.obs-mip.fr}
}

\offprints{F. Ligni\`eres}

%\mail{Francois.Lignieres@bagn.obs-mip.fr}

\date{Received  \today / Accepted ??}

%\authorrunning{F. Ligni\`eres and M. Rieutord}
%\titlerunning{Acoustic oscillations in centrifugally flattened polytropic star}

\abstract
{}
{A new non-perturbative method to compute accurate oscillation modes in rapidly rotating stars is presented.}
{In this paper, the effect of
the centrifugal force is fully taken
into account while the Coriolis force is neglected. This assumption is valid when the time scale of the oscillation
is much shorter than the inverse of the rotation rate and is expected to be suitable
for high
radial order p-modes of $\delta$ Scuti stars.
Axisymmetric p-modes have been computed in uniformly rotating polytropic models of stars.}
{In the frequency and rotation range considered, we found that as rotation increases
(i) the asymptotic structure of the non-rotating frequency spectrum is first destroyed then
replaced by a new form of organization 
(ii) the mode amplitude
tends to concentrate
near the equator (iii) differences with perturbative
methods become significant as soon as the rotation rate exceeds about fifteen percent
of the Keplerian limit.
The implications for the seismology of rapidly rotating stars are then discussed.}
{}


\maketitle

\section{Introduction}

Since helioseismology revolutionized our knowledge of the solar interior, great advances in stellar structure and evolution
theory 
are expected from asteroseismology.
Major efforts including space missions are under way to detect pulsation frequencies with unprecedented accuracy
across the HR diagram \citep{Ca95,Wa03}. 
To draw information from the observed frequencies,
seismology relies on the theoretical computation
of eigenmodes for a given model of star.
Yet, except for slowly rotating stars, the effect of rotation on the gravito-acoustic modes is 
not fully taken into account in the present theoretical calculations (e.g. Rieutord 2001).

Rotational effects have been mostly studied through perturbative methods.
In this framework, both $\Omega/\omega$, the ratio of the rotation rate $\Omega$ to the mode frequency $\omega$, 
and $\Omega / \sqrt{GM/R^3}$, the square root of the ratio of the centrifugal force to the gravity at equator
are assumed to be small and of the same order.
Solutions valid up to the first, second, and even third order in $\Omega/\omega$
have been obtained respectively  by \citet{Le51}, \citet{Sa81} and \citet{Go98}.
The first order analysis proved fully adequate to match the observed acoustic frequency of the slowly rotating sun
\citep{Dz92}.
At the other extreme, the perturbative methods are not expected to be correct for stars approaching
the Keplerian limit $\Omega_K = \sqrt{GM/R_e^3}$, where $R_e$ is the equatorial radius. 
Achernar is a spectacular example of such star since interferometric observations showed
a very important distortion of its surface, the equatorial radius $R_e$ being
at least one and a half time larger than 
the polar radius $R_p$ \citep{Do03}. In the context of Roche models, such a flattening occurs at the Keplerian limit $\Omega_K$.
For intermediate rotation rates, second or third order perturbative methods might be used, but
the main problem is that the limit of validity of the perturbative methods is unknown.
Departures from the perturbative results would impact the
values of individual frequency but also other
properties which are commonly used to analyze the
spectrum of observed frequencies. This concerns in particular
the rotational splitting,
the asymptotic large and small frequency separations or
the mode visibility.

New methods able to compute accurate eigenmodes in rotating stars
are therefore needed to allow progress in the seismology 
of rapidly rotating stars. Incidentally, such methods would also assess the limit of validity
of perturbative analysis.
The main difficulty comes from the fact that, except in the special cases of spherically symmetric media
and uniform density ellipsoids,
the eigenvalue problem of
gravito-acoustic resonances in arbitrary axially symmetric cavities
is not separable in the radial and meridional variables.
For self-gravitating and rotating stars, a two-dimensional eigenvalue problem has to be solved.

\citet{Cl81,Cl98} made the first attempts to solve this eigenvalue problem for gravito-acoustic modes, investigating
various numerical schemes. However, the accuracy of his calculations is generally difficult to estimate.
Moreover, the different numerical schemes could not converge for low frequency g-modes when $\Omega/\omega$ exceeds about 0.5.
Since then, eigenmodes in this frequency range
have been successfully calculated by \citet{Di00} using spectral
methods. These authors however did not consider the effect of the centrifugal acceleration in their model.
The search for unstable modes in neutron stars also triggered the development of
numerical schemes able to solve the two-dimensional eigenvalue problem.
But only surface gravity modes (f-modes) and some inertial modes (r-modes)
have been determined in this context \citep{Yo05}.
\citet{Es04} reported calculations of adiabatic acoustic modes in MacLaurin spheroids of uniform density neglecting
the Coriolis force, the potential perturbation and the Brunt-V\"{a}is\"{a}l\"{a} frequency.

In this paper, we present a new method to compute accurate eigenmodes in rotating stars.
For the first application of the method, we only consider the effect of
the centrifugal force through its impact on the equilibrium state of the star;
we thus neglect the Coriolis force. This assumption is valid when the time scale of the oscillation
is much shorter than the inverse of the rotation rate and is expected to be suitable
for high
radial order p-modes of $\delta$ Scuti stars.
The problem is further simplified by using uniformly rotating polytropes
as equilibrium models and assuming adiabatic perturbations as
well as the Cowling approximation.
Low frequency axisymmetric p-modes have been computed for rotation rates
varying from $\Omega = 0$ up to $\Omega / \Omega_K = 0.59 $, this ratio
corresponding
to a typical $\delta$ Scuti star ($M=1.8 M_{\sun}, R=2 R_{\sun}$) with an equatorial velocity of $240$ \kms.
The centrifugal force modifies the effective gravity in two ways: it makes it smaller and aspherical.
Decreasing the effective gravity should affect sound waves by reducing the sound speed
inside the star and by increasing the star's volume, thus potentially the volume of the resonant cavity.
Besides,
the physical consequences of the non-spherical geometry are basically unknown.

In the following, the formalism and the numerical method are presented
along with accuracy tests.
Then,
the parameter range of the calculations is given together with
the method used to label the eigenmodes. 
The structure of the frequency spectrum, some
properties of the eigenfunctions and 
the differences with perturbative methods 
are further
analyzed as a function of the rotation rate.
These results are discussed in the last section.

\section{Formalism} 

Accurate numerical solutions of 2D eigenvalue problems require a careful choice
of the numerical method and the mathematical formalism.
In this section we explain the choices that have been made for the variables, the coordinate system, the numerical discretization,
and the method to solve the resulting algebraic eigenvalue problem.
All play a role in the accuracy of the eigenfrequency determinations that will be presented at the end of
this
section.

\subsection{Equilibrium model}

The equilibrium model is
a self-gravitating uniformly rotating polytrope.
It is therefore governed by a polytropic relation, the hydrostatic equilibrium in a rotating frame,
and Poisson's equation for the gravitational potential:
\begin{eqnarray} \label{hyd}
P_0 &=& K \rho_0^{1+1/N} \\
0 &=& -\nab P_0 - \rho_0 \nab \left( \psi_0 -\Omega^2 s^2/2 \right)  \\
\lapl \psi_0 &=& 4\pi G\rho_0
\end{eqnarray}
where $P_0$ is the pressure, $\rho_0$ the density, $K$ the polytropic constant,
$N$ the polytropic index, $\psi_0$ the gravitational potential,
$s$ the distance to the
rotation axis and $G$ the gravitational constant.  

The polytropic relation and uniform rotation ensure that the
fluid is barotropic.
A pseudo-enthalpy can then be introduced $h_0=\int dP_0/\rho_0 = (1+N)P_0/\rho_0$ 
and the integration of the hydrostatic equation reads:
\begin{equation}
\label{eq:hydrostatic.enthalpy}
h_0 = h_c -(\psi_0 - \psi_c) + \frac{1}{2}\Omega^2 s^2
\end{equation}
where the subscript ``$c$'' denotes the value in the center of the polytropic model.
Equation \eq{eq:hydrostatic.enthalpy} is then inserted into Poisson's equation to yield:
\begin{equation}
\label{eq:Poisson.phi}
\lapl \psi_o = 4 \pi G \rho_c \left( 1 - \frac{\psi_o - \psi_c}{h_c} +
\frac{\Omega^2 s^2}{2 h_c} \right)^{N}
\end{equation}

Equation \eq{eq:Poisson.phi} is solved numerically, using an iterative scheme.
Since the shape of the star is not spherical, a system of coordinates
$(\zeta,\theta,\phi)$ based on \citet{Bo98} is used, such that $\zeta=1$
corresponds to the surface of the spheroid (more details on the coordinate system are given in section 2.3).
This enables the use of spectral
methods both for the radial coordinate $\zeta$ and the angular ones.  
The numerical method is further detailed in 
Rieutord et al. (2005).

\subsection{Perturbation equations and boundary conditions}

Neglecting the Coriolis force, the linear equations governing the evolution
of small amplitude adiabatic perturbations read:

\begin{equation} \label{div}
{\partial}_t \rho + \nab \cdot (\rho_{0} \vi)=0,
\end{equation}    

\begin{equation} \label{vel}
\rho_{0} {\partial}_t \vi =
- \nab P + \rho \vec{g}_0 - \rho_{0} \nab \psi, 
\end{equation}  

\begin{equation} \label{adia}
{\partial}_t P + \vi \cdot \nab P_0 = c_0^{2} \lp {\partial}_t \rho + \vi \cdot \nab \rho_0 \rp, 
\end{equation}

\begin{equation} \label{poisson}
\Delta \psi = 4 \pi G \rho
\end{equation} 
where $\vi$, $\rho$, $P$ and $\psi$, are the perturbations of velocity,
density, pressure and gravitational potential. The sound speed is 
$c_0 = \sqrt{{\Gamma}_{1,0} P_0 / \rho_{0}}$, ${\Gamma}_{1,0}$ being the first adiabatic exponent of the gas,
and the effective gravity $\vec{g}_0 = - \nab \left( \psi_0 -\Omega^2 s^2/2 \right)$ has been introduced.
In the framework of polytropic models of stars, the polytropic relation \eq{hyd} is assumed to give a reasonably
good approximation
of the relation between the pressure and the density of the equilibrium state.
As the first adiabatic exponent ${\Gamma}_{1,0}$ is obtained from the
equation of state of the gas,
${\Gamma}_{1,0}$ is in general 
not equal to  $1+1/N$.

We simplified Eqs. \eq{div}, \eq{vel}, \eq{adia}, \eq{poisson} following two constraints:
First,
the governing equations should be written for general coordinate systems
because we shall use a surface-fitting non-orthogonal coordinate system.
Second, they should take the form $\cal{M} \vec{X} = \lambda \cal{Q} \vec{X}$ where $\lambda$ is the eigenvalue, 
$\vec{X}$ is the eigenfunction, and 
$\cal{M}$ and $\cal{Q}$ are
linear differential operators. Indeed, 
the method that we shall use to solve the algebraic eigenvalue problem obtained after discretization 
works for problem reading
$[M] X = \lambda [Q] X$, where $X$ is the discretized eigenvector and, $[M]$ and $[Q]$ are matrices.
Taking the time derivative of Eqs. \eq{vel}
and \eq{poisson} and using Eqs. \eq{div} and \eq{adia} to eliminate
the pressure and density perturbations, we obtain two equations for the
velocity and the gravitational potential perturbation:

\begin{equation} \label{adib}
{\partial}^{2}_{tt} \vi = \nab \lp c_0^2 \chi + \vi \cdot \vec{g}_0 - {\partial}_t \psi \rp
- \chi \vec{A}_0
\end{equation}  

\begin{equation} \label{poissonbis}
\Delta {\partial}_t \psi = - 4 \pi G \lp\vi \cdot \nab {\rho}_0  +
{\rho}_0 \chi \rp
\end{equation}
where $\chi = \nab \cdot \vi$ is the divergence of the velocity. The vector
$\vec{A}_0$ characterizes the stratification of the equilibrium model:
\begin{equation} \label{A0}
 \vec{A}_0 = c^2_0 \lp \frac{1}{{\Gamma}_{1,0}}\frac{\nab P_0}{P_0} - 
\frac{\nab \rho_{0}}{\rho_{0}} \rp = 
\frac{c_0^2 N_0^2}{\|\vec{g}_0\|} \vec{n}_0, \\
\end{equation}
where
$N_0$ is the Brunt-V\"{a}is\"{a}l\"{a} frequency and
$\vec{n}_0$ is the unit vector in the direction opposite to
the effective gravity defined by:
\begin{equation} \label{g0}
\vec{g}_0 = - \|\vec{g}_0\| \vec{n}_0.
\end{equation}
Note that the barotropicity
of the fluid has been used to obtain \eq{adib}.

In addition to the gravitational potential perturbation, the right hand sides of
Eqs. \eq{adib} and \eq{poissonbis} only involve the divergence of
the velocity and the scalar product of the velocity with vectors parallel
to gravity.  Then, the scalar product of Eq. \eq{adib} with gravity,

\begin{equation} \label{oneok}
{\partial}^2_{tt} \vi \cdot \vec{g}_0 = \vec{g}_0 \cdot \nab \lp c_0^2
\chi + \vi \cdot \vec{g}_0 -
{\partial}_t \psi \rp - \chi \vec{g}_0 \cdot \vec{A}_0
\end{equation}
and the divergence of Eq. \eq{adib},

\begin{equation} \label{twofirst}
{\partial}^2_{tt} \chi = \Delta (c_0^2 \chi + \vi \cdot \vec{g}_0 -
\partial_t \psi) - \nab \cdot (\chi \vec{A}_0)
\end{equation} 
yield, together with Eq. \eq{poissonbis}, a complete set of
differential equations for the variables $\vi \cdot \vec{g}_0$, $\chi$
and $\psi$.  \citet{Pe38} who studied the oscillations of spherically
symmetric polytropes considered similar variables but, instead
of Eq. \eq{adib}, used a combination of Eqs. \eq{oneok} and
\eq{adib} which does not involve second order derivative with respect to the
radial coordinate.  For general system of coordinate as well, the order
of the differential system can be lowered replacing Eq. \eq{adib}
by the following one:

\beqan
{\partial}^2_{tt} \lc \chi - g^{11} {\partial}_1 \!\! \lp \frac{\vi \cdot \vec{g}_0}{g_0^1} \rp \rc =
{\cal L}(c_0^2 \chi + \vi \cdot \vec{g}_0 - \partial_t \psi) - \nonumber \\
\nab\cdot(\chi \vec{A}_0) + 
g^{11} {\partial}_1 \!\! \lp \frac{\chi \vec{g}_0 \cdot \vec{A}_0 }{g_0^1} \rp
\eeqan{twook}
where the linear operator ${\cal L}$, defined by,

\begin{equation}
{\cal L}(F) = \Delta F - g^{11} {\partial}_1 \!\! \lp\frac{\vec{g}_0 \cdot \nab F}{g_0^1} \rp
\end{equation}
does not contain second order derivatives with respect to the first
coordinate $x^1$.  In this expression, $g_0^1$ is the first contravariant
component of the gravity in the natural basis 
$(\vec{E}_1, \vec{E}_2, \vec{E}_3)$ defined by 
$\vec{E}_i = \partial \vec{OM} / \partial x^i$, and $g^{11}$ is the
first contravariant component of the metric tensor.

The equations are non-dimensionalized using the equatorial radius, $R_e$,
as length unit, the density at the center of the polytrope, $\rho_c$,
as density unit and $T_0 = \lp 4 \pi G \rho_c \rp^{-1/2}$ as time unit.
As we seek for harmonic solutions in time, the variable are written $F =
\hat{F} \exp(i \omega t)$. Dropping the hat and denoting dimensionless
quantities as previous dimensional ones, the governing equations
read:

\beqan
&&\lambda W  = \vec{g}_0 \cdot \nab \lp c_0^2 \chi + W +
\Psi \rp - c_0^2 N_0^2 \chi \label{one} \\
&&\lambda \lc \chi - g^{11} {\partial}_1 \!\! \lp \frac{W}{g_0^1} \rp \rc
= \nonumber \\
&&{\cal L}(c_0^2 \chi + W + \Psi ) - \nab \cdot (\chi \vec{A}_0) +
g^{11} {\partial}_1 \!\! \lp \frac{c_0^2 N_0^2 \chi}{g_0^1} \rp
\label{two} \\
&&0 = \Delta \Psi - d_0 W - {\rho}_0 \chi 
\eeqan{three}
\noi
where $\lambda = - \omega^2$, $W = \vi \cdot \vec{g}_0$, $\Psi = - i \omega
\psi$ and $d_0$ denotes

\begin{equation} 
d_0 = \frac{\|\nab \rho_0\|}{\|\vec{g}_0\|}
\end{equation}

\noi Another form of these equations may be obtained replacing $W$ by a new 
variable $U = c_0^2 \chi + W + \Psi$. The set of equations then reads:

\begin{equation} \label{oneII}
\vec{g}_0 \cdot \nab U - c_0^2 N_0^2 \chi =
\lambda (U - \Psi - c_0^2 \chi)
\end{equation}
 
\beqan
{\cal L}(U) - \nab \cdot (\chi \vec{A}_0) +
g^{11} {\partial}_1 \!\! \lp \frac{ c_0^2 N_0^2 \chi }{g_0^1} \rp
= \nonumber \\
\lambda \lc - g^{11} {\partial}_1 \!\! \lp \frac{U-\Psi}{g_0^1} \rp
+ \chi + g^{11} {\partial}_1 \!\! \lp \frac{c_0^2 \chi}{g_0^1} \rp \rc
\eeqan{twoII}
 
\begin{equation} \label{threeII}
- d_0 U + (d_0 c^2_0 - {\rho}_0) \chi + d_0 \Psi + \Delta \Psi = 0
\end{equation}
\noi
As in \citet{Pe38}, the boundary conditions are that the gravitational potential vanishes
at infinity and that $U$ and $\chi$ be regular everywhere.

\subsection{Coordinates choice}

The choice of the coordinate system has been guided by
two considerations.
First, for the accuracy of the numerical method, it seems preferable to apply the
boundary conditions on a surface of coordinate.  This imposes that 
the stellar surface is described by 
an equation
$\zeta = \mbox{constant}$, where $\zeta$ is one of the coordinates. Second,
when using spherical harmonic expansions, the regularity conditions at the center
have a simple form for spherical coordinates only.
Therefore, 
the coordinate system should become spherical near the center.
If $(r, \theta, \phi)$ denotes the usual spherical
coordinates and $r= S(\theta)$ describes the surface,
families of coordinates
$(\zeta,\theta',\phi')$ verifying both conditions have been proposed by \cite{Bo98}:

\begin{equation}\label{eq:coor}
\left\{\begin{array}{l}
r  = r(\zeta, \theta') \\
\theta = \theta' \\
\phi = \phi', \end{array}\right.
\end{equation}
where
\begin{equation}\label{eq:dep}
r(\zeta, \theta) = \alpha \zeta + A(\zeta) \lc S(\theta) - \alpha \rc
\end{equation}
The polynomial $A(\zeta) = (5 \zeta^3 - 3 \zeta^5)/2$ 
ensures that $r \propto \zeta$ near the center
and that $\zeta = 1$ describes the surface $r= S(\theta)$.
The scalar $\alpha$ is
chosen
so that the transformation $(r, \theta, \phi) \mapsto (\zeta, \theta,
\phi)$ is not singular and the numerical convergence is fast.  We find that $\alpha = 1 - \epsilon$ is
a convenient choice, $\epsilon = 1 - R_p/R_e $ being the flatness
of the star surface. In the following, we shall refer to $\zeta$ as the pseudo-radial coordinate.

To express the governing equations in this non-orthogonal coordinate system, we
use the covariant and contravariant components of the corresponding metric
tensor. The non-vanishing components read:

\begin{equation} \label{eq:metr}
\begin{array}{ll}
g_{11} = r_{\zeta}^{2} & g_{12} = g_{21} = r_{\zeta} r_{\theta} \\
g_{22} = r^{2} + r_{\theta}^{2} & g_{33} = r^{2} \sin^2\!\theta \\
g^{11} = (r^{2} + r_{\theta}^{2})/(r^{2} r_{\zeta}^{2}) & g^{12} = g^{21} = - r_{\theta}/(r^{2} r_{\zeta})\\
g^{22} = 1/r^{2} & g^{33} = 1/(r^{2} \sin^2\!\theta),
\end{array}
\end{equation}
\noindent and the square-root of the absolute value of the metric tensor determinant is:
\begin{equation} \label{eq:det}
\sqrt{\mid g \mid} = r^{2} r_{\zeta} \sin\! \theta.
\end{equation}

In Appendix A, the linear operators involved in Eqs. \eq{oneII},
\eq{twoII}, \eq{threeII} are expressed
in terms of the partial derivatives of $r(\zeta,\theta)$.
Note that, for vectorial operators, we used the natural basis defined above.

\subsection{The numerical method}

The method follows and generalizes the one 
presented in \citet{Ri97}.
The numerical discretization is done with spectral methods,
spherical harmonics for the angular coordinates $\theta$ and $\phi$
and Chebyshev polynomials for the pseudo-radial coordinate $\zeta$.
The variables $U$, $\Psi$ and $\chi$ are expanded
into spherical harmonics:
\begin{equation} 
U(\zeta,\theta,\phi) = \sum_{\ell=0}^{L}\sum_{m=-\ell}^{+\ell} U^\ell_m (\zeta)
\YL(\theta,\phi).
\end{equation} 
and equivalent expressions for $\Psi$ and $\chi$, where $\ell$ and $m$ are respectively the degree and the
azimuthal number of the spherical harmonic $\YL(\theta,\phi)$. Then, the governing equations are projected onto spherical harmonics
to obtain
a system of ordinary differential equations (ODE) of the variable $\zeta$ for
the coefficients of the spherical harmonic expansion $U^{\ell}_m(\zeta), \chi^{\ell}_m(\zeta), \Psi^{\ell}_m(\zeta)$.
This system is then discretized on the collocation points of a Gauss-Lobatto grid associated with Chebyshev polynomials.
It results in an algebraic eigenvalue problem
$[M] X = \lambda [Q] X$, where $X$ is the eigenvector of $L \times N_r$ components and $[M]$ and $[Q]$ are square matrices of $L \times N_r$ lines,
$L$ and $N_r$ being
respectively the truncation
orders on the
spherical harmonics and Chebyshev basis.
The algebraic eigenvalue problem is solved  
using either
a QZ algorithm or an Arnoldi-Chebyshev algorithm.
The QZ algorithm provides the whole spectrum of eigenvalues while the iterative calculation based
on the Arnoldi-Chebyshev algorithm computes a few eigenvalues around a given value of the frequency.

Because of the symmetries of the equilibrium model with respect to the rotation axis and
the equator, one obtains a separated eigenvalue problem
for each absolute value of the azimuthal number $\mid \! m \! \mid$
and each parity with respect to the equator.
Thus, for a given $m \geq 0$, we have two independent sets of ODE coupling
the coefficients of the spherical harmonic expansion  having respectively
even and odd degree numbers, that is:
\begin{equation}
\begin{array}{lll}
\hat{U}^{+} = U^{m+2k}_m (\zeta) & \hat{\chi}^{+} = \chi^{m+2k}_m (\zeta) & \hat{\Psi}^{+}=\Psi^{m+2k}_{m}
(\zeta)\\
\hat{U}^{-} = U^{m+2k+1}_m (\zeta) & \hat{\chi}^{-} = \chi^{m+2k+1}_{m} (\zeta) & \hat{\Psi}^{-}=\Psi^{m+2k+1}_m
(\zeta),
\end{array}
\end{equation}
\noindent where $0 \leq k < +\infty$.
% $\hat{U}^{+} = U^{m+2k}_m (\zeta),
%\; \hat{\chi}^{+} = \chi^{m+2k}_m (\zeta), \; \hat{\Psi}^{+}=\Psi^{m+2k}_{m}
%(\zeta), \; 0 \leq k < +\infty$ and $\hat{U}^{-} = U^{m+2k+1}_m (\zeta), \;
%\hat{\chi}^{-} = \chi^{m+2k+1}_{m} (\zeta), \; \hat{\Psi}^{-}=\Psi^{m+2k+1}_m
%(\zeta), \; 0 \leq k < +\infty$.  
The solutions of these two ODE systems are 
either symmetric
or antisymmetric with respect to the equator.

The two sets of ODE can be written in the form:

\begin{equation} \label{form}
\lp {\cal A} \ddnz{} + {\cal B} \dnz{} + {\cal C} \; \mbox{Id} \rp \vec{\Xi} = 
\lambda \lp {\cal D} \dnz{} + {\cal E} \rp \vec{\Xi},
\end{equation}  
\noi where 
$\vec{\Xi}$ denotes
\begin{eqnarray}
\vec{\Xi}^{+} = \left| \begin{array}{l}
\hat{U}^{+} \\
\hat{\chi}^{+} \\
\hat{\Psi}^{+} \\
\end{array} \right. & \ltex{or}  &
\vec{\Xi}^{-} = \left| \begin{array}{l}
\hat{U}^{-} \\
\hat{\chi}^{-} \\
\hat{\Psi}^{-} \\
\end{array} \right.
\end{eqnarray}
\noi       
and where the matrices are defined by blocks as follows:
\begin{equation}
 {\cal A} = \lp \begin{array}{ccc}
       0 & 0 & 0 \\
       0 & 0 & 0 \\
       0 & 0 & \mbox{A}_{33} \end{array} \rp
\qquad
{\cal B} = \lp \begin{array}{ccc}
       \mbox{B}_{11} & 0 & 0 \\
       \mbox{B}_{21} & \mbox{B}_{22} & 0 \\
       0 & 0 & \mbox{B}_{33}  \end{array} \rp
\end{equation}

\begin{equation}
{\cal C} = \lp \begin{array}{ccc}
       \mbox{C}_{11} & \mbox{C}_{12} & 0 \\
       \mbox{C}_{21} & \mbox{C}_{22}  & 0 \\
       \mbox{C}_{31} & \mbox{C}_{32} & \mbox{C}_{33}  \end{array} \rp
\end{equation} 

\begin{equation}
 {\cal D} = \lp \begin{array}{ccc}
       0 & 0 & 0 \\
       \mbox{D}_{21} & \mbox{D}_{22} & -\mbox{D}_{21}  \\
       0 & 0 & 0  \end{array} \rp
\qquad 
 {\cal E} = \lp \begin{array}{ccc}
       \mbox{E}_{11} & \mbox{E}_{12} & -\mbox{E}_{11} \\
       \mbox{E}_{21} & \mbox{E}_{22} & -\mbox{E}_{21}  \\
       0 & 0 & 0  \end{array} \rp
\end{equation}

\noi Equivalently, one can write Eq. \eq{form} as:
\greq
\label{eq:sys}
\lp \mbox{B}_{11} \dnz{} + \mbox{C}_{11} \rp \hat{U} + \mbox{C}_{12}
\hat{\chi} =
\lambda \lc \mbox{E}_{11} \lp \hat{U} - \hat{\Psi} \rp + \mbox{E}_{12}
\hat{\chi} \rc \\
\lp \mbox{B}_{21} \dnz{} + \mbox{C}_{21} \rp \hat{U} + \lp \mbox{B}_{22}
\dnz{} + \mbox{C}_{22} \rp \hat{\chi} = \\
\lambda \lc \lp \mbox{D}_{21} \dnz{} + \mbox{E}_{21} \rp \lp \hat{U} - \hat{\Psi} \rp + 
\lp \mbox{D}_{22} \dnz{} + \mbox{E}_{22} \rp \hat{\chi} \rc \\ 
\mbox{C}_{31} \hat{U} + \mbox{C}_{32} \hat{\Psi} + \lp \mbox{A}_{33} \ddnz{} + \mbox{B}_{33} \dnz{} + \mbox{C}_{33} \rp \hat{\Psi} = 0
\egreq 

\noi Each sub-matrices can be expressed in terms of the two following functionals:
\begin{equation}\label{eq:ii}
\gi (\mbox{G})= 2\pi \int_0^{\pi} \mbox{G}(\zeta,\theta) \YTL(\theta) \YTLP(\theta) \sin \theta d \theta
\end{equation}
\begin{equation} \label{eq:jj}
\gj (\mbox{G}) = 2\pi \int_0^{\pi} \mbox{G}(\zeta,\theta) \YTL(\theta) \DYTLP(\theta) \sin \theta d \theta
\end{equation}
\noi where $\YTL(\theta) = \YL(\theta,\phi) e^{-im\phi}$ is a normalized Legendre polynomial.

In Appendix B, all the coefficient of the sub-matrices are made explicit in terms
of the function $r(\zeta,\theta)$ and its first and second order derivatives as well as in terms of 
the enthalpy of the equilibrium model, its first and second order
derivatives.

In the following, we consider the Cowling approximation thus neglecting the gravitational potential perturbation.
The ODE system \eq{form} is simplified accordingly and in particular reduces to the first order.

\section{Tests and accuracy}

The formalism and the numerical method presented in the previous section have been tested
and the accuracy of the frequency determinations has been estimated.

\subsection{Tests}

A first test of the method has been performed in the case of axisymmetric ellipsoids 
of uniform density.
We choose this configuration because the eigenvalue problem is fully separable using the
oblate spheroidal coordinates $(\xi, \eta, \phi)$
defined as $(x = a \cosh \xi \sin \eta \sin \phi, \; y = a \cosh \xi \sin \eta \cos \phi,
\; z = a \sinh \xi \cos \eta)$,
where $0\leq \xi <+ \infty$, $0\leq \eta \leq \pi$ et $0\leq \phi \leq 2
\pi$. The eigenfrequencies obtained with this method were compared with the eigenfrequencies
computed with our general method, $S(\theta)$ describing an ellipse.
We found the same frequencies with a high degree of accuracy for arbitrary values of the ellipsoid flatness
between $0$ and $0.5$. Moreover, as the flatness goes to zero, the frequencies
were found to converge towards the values given by a first order perturbative analysis in terms
of
flatness. More details about this test are given in \citet{Li01} and \citet{Li04}.

The frequencies of axisymmetric p-modes in a self-gravitating uniformly rotating $N=3$ polytrope
that will be presented in the following sections have been also tested.
As shown in the previous section, the method involves lengthy analytical calculations 
of the
coefficients of the ODE system \eq{form}.
Terms involved in the non-rotating case have been tested by comparing our result
with the p-modes frequencies in a non-rotating self-gravitating $N=3$ polytrope published in \citep{Da94}.
The relative error is smaller than $10^{-7}$ for the $\ell=0$ to $3$, $n=1$ to $10$ modes.
In the rotating case, we compared our results with the ones obtained by 
solving the same problem
but using a different form of the starting
equations. This alternative system of equations aims at including the Coriolis force;
thus the variables and the resulting ODE systems to be solved are
different. 
We verified that when the terms involving the Coriolis force are omitted
from the equations,
eigenfrequencies presented in the next section are recovered with a very high precision \citep{Re06}.
For instance, the maximum relative error on the eigenfrequency for all the frequencies computed at $\Omega/\Omega_K = 0.59 $ has been found
of the order of $10^{-6}$, for a given set of numerical parameters.
This last test gives us strong confidence in
the method and its implementation.

\begin{figure}[h]
\resizebox{\hsize}{!}{\includegraphics{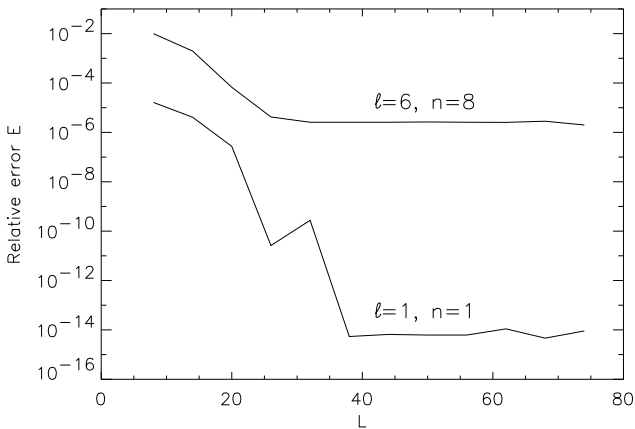}}
\caption{Evolution of the frequency relative error, $E(L) = |\omega(L) - \omega(L^{\mathrm{max}})| / \omega(L^{\mathrm{max}})$, as the spatial resolution
 in latitude is increased. Two modes
labeled ($\ell=1,\;n=1$) and
($\ell=6,\;n=8$) are considered at the rotation rate $\Omega/\Omega_K = 0.46 $ with
$L^{\mathrm{max}} = 80$ and $N_r = 61$.}
\label{er_l}
\end{figure}

\subsection{Accuracy}

As pointed out by \citet{Cl81,Cl98}, accurate numerical solutions of the 2D eigenvalue problem
are not easy to obtain. It is therefore important to estimate the accuracy of our method.
In the following we first investigate the influence of the spatial resolution on the
eigenfrequencies and then consider other sources of errors.

\begin{figure}[htb]
\resizebox{\hsize}{!}{\includegraphics{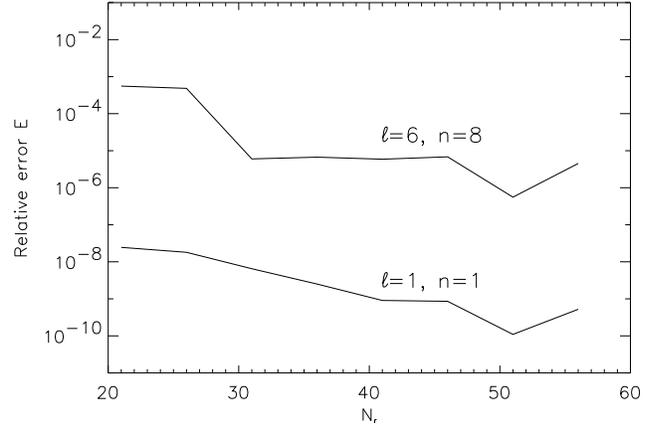}}
\caption{Evolution of the frequency relative error,
$E(N_r) = |\omega(N_r) - \omega(N_r^{\mathrm{max}})| / \omega(N_r^{\mathrm{max}})$
as the resolution in the radial coordinate is increased. Two modes
labeled ($\ell=1,\;n=1$) and
($\ell=6,\;n=8$) are considered at a rotation rate $\Omega/\Omega_K = 0.46 $ with
$N_r^{\mathrm{max}} = 61$ and $L = 62$.}
\label{er_r}
\end{figure}

A relative spectral error $E$ is defined as the absolute value of the relative difference 
between the frequency computed at a given resolution and the frequency obtained
at the maximum resolution considered. Let us first consider the effects of the angular resolution.
Fig.~\ref{er_l} displays $E(L) = |\omega(L) - \omega(L^{\mathrm{max}})| / \omega(L^{\mathrm{max}})$ as 
a function of $L$, the truncation order of
the spherical harmonic expansion, for two axisymmetric modes labeled ($\ell=1,\;n=1$) and
($\ell=6,\;n=8$) whose spatial structures are dominated by large and small length scales,
respectively (the labeling of the mode will be described in the next section). 
The maximum angular resolution is $L^{\mathrm{max}} = 80$, the resolution in the pseudo-radial coordinate
is fixed to $N_r = 61$ and the rotation rate is $\Omega/\Omega_K = 0.46$.
In the same way, Fig.~\ref{er_r} illustrates the effects of the pseudo-radial resolution by showing 
$E(N_r) = |\omega(N_r) - \omega(N_r^{\mathrm{max}})| / \omega(N_r^{\mathrm{max}})$ as
a function of $N_r$, the truncation order of
the Chebyshev polynomial expansion, for the same modes and rotation rate. The maximum radial resolution is 
$N_r^{\mathrm{max}} = 
61$ and the latitudinal resolution is fixed to $L=62$.
In both figures, we observe that the error first decreases and then reaches a plateau which means 
that a better approximation
of the eigenfrequency cannot be obtained by increasing the spatial resolution.
The plateau are significantly higher for the ($\ell=6,\;n=8$) mode than for the ($\ell=1,\;n=1$) mode.
We verified that this difference is due to the presence of smaller radial length scales
(rather than to smaller angular length scales).

Even when the spatial resolution is sufficient, two other sources of numerical errors can indeed limit
the accuracy of eigenfrequency determination.
First, the component of the matrix $L$ and $M$ being
computed numerically they are determined with a certain error.
Second, even when this error is reduced to round-off errors,
the accuracy of the algebraic eigenvalue solver, the Arnoldi-Chebyshev algorithm, remains limited.

\begin{figure}[htb]
\resizebox{\hsize}{!}{\includegraphics{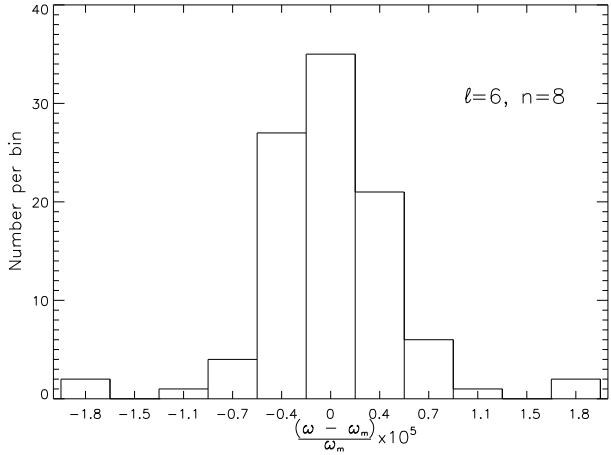}}
\caption{Histogram of 100 frequencies obtained for 100 different values of the initial guess of the Arnoldi-Chebyshev algorithm
randomly chosen in a small interval around $\omega_m = 12.0547$. The standard deviation of this frequency distribution
$\sigma= 5.6 \times 10^{-6}$ measures the algorithm relative error on the frequency for this particular mode labeled ($\ell=6,\;n=8$). 
The spatial resolution is $N_r = 61$ and $L = 62$ and the rotation rate is $\Omega/\Omega_K = 0.46$.}
\label{er_ss}
\end{figure}

Errors on the matrix component that arise from quadratures
(see Eqs. \eq{eq:ii} and \eq{eq:jj}) can approach round-off errors
using weighted Gauss-Lobatto quadratures. The other source of error
in the matrix components comes from the computation of equilibrium quantities.
Indeed, the accuracy of the enthalpy, its first and second  derivatives 
and the surface shape,
is at best limited by the effect of round-off errors on the convergence of the algorithm
used to compute the polytropic stellar models.
The effect of these errors on the eigenfrequencies have been investigated and
appears to be smaller than the effect of the Arnoldi-Chebyshev algorithm itself
which is now described.

As any solver in linear algebra, the Arnoldi-Chebyshev algorithm amplifies the round-off error that
affect the matrix components. Thus, the error on the eigenvalue and the associated eigenvector is usually
much larger than the round-off error of double precision arithmetic. The accuracy of the Arnoldi-Chebyshev
algorithm has been studied
in details by \citet{Va00}
in the context of inertial modes in a spherical
shell where the matrix component are known analytically.
Theoretically, it can be estimated by
computing the spectral portrait of the
eigenvalue problem $[M] X = \lambda [Q] X$, which shows how small variations of $[M]$ and $[Q]$
affects the determination of each eigenfrequencies.
In fact, as the iterative Arnoldi-Chebyshev algorithm requires an initial guess of the eigenfrequency,
a practical alternative to measure the accuracy of a frequency
determination is to compute frequencies for slightly different values of the initial guess.
This has been done for a large number (100) of initial guess values randomly distributed around the eigenfrequency 
of the ($\ell=1,\;n=1$) and
($\ell=6,\;n=8$) modes.
The histogram in Fig.~\ref{er_ss} shows the resulting frequencies distribution around a most probable mean
eigenfrequency. The width of the histogram determined by the standard deviation of the distribution provides a measure
of the algorithm accuracy. The standard deviation $\sigma$ is equal to $5.6 \times 10^{-6}$ for the ($\ell=6,\;n=8$) mode and to $6.2 \times 10^{-10}$
for the ($\ell=1,\;n=1$) mode.
The error thus grows with the radial order of the mode, this trend being general in our results (as in Valdettaro et al., 2000).
Moreover, the width of the histogram does not depend on the amplitude of the initial guess perturbation provided
it is sufficiently small.

We also observe that, except for the dependence of the ($\ell=1,\;n=1$) frequency on the
angular resolution, the levels of the plateau  shown in Figs.~\ref{er_l} and ~\ref{er_r} are of the same order as the 
error of the algorithm.
It means that, in these cases, the changes in the matrix component and size associated with
the modification of the resolution
have a similar effect on the frequency as varying the initial guess of the algorithm.
However, the convergence of the ($\ell=1,\;n=1$) frequency at a $10^{-14}$ level, much lower than the $6.2 \times 10^{-10}$ accuracy of the algorithm,
shows that it is not always true and that the spectral error can underestimate the true error.

Although it is too demanding to compute a global accuracy 
by repeating the statistical study on the initial guess for all eigenfrequencies,
the relative accuracy on all tested frequencies is
always better than $2 \times 10^{-5}$ using double precision arithmetic.

Note that another potential source of error will be discussed below when describing
avoided crossings between modes.

\begin{figure*}[htb]
\includegraphics{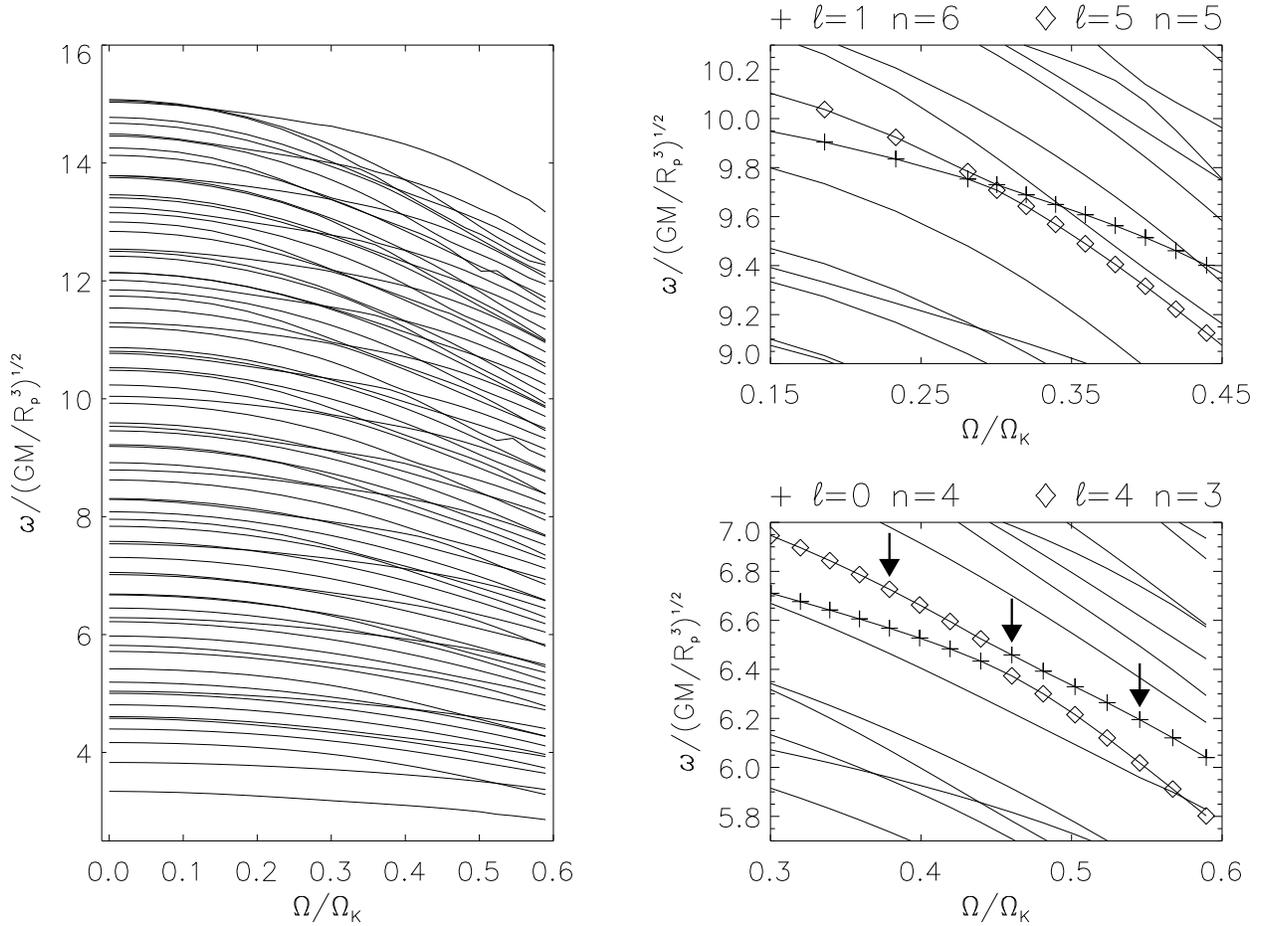}
\caption{
Evolution of
all the computed p-modes frequencies from $\Omega=0$ to $\Omega / \Omega_K = 0.59$. 
The frequencies have been adimensionalized by $(GM/R_p^3)^{1/2}$ because we expect that the polar radius $R_p$ does
not change much
as the rotation of the star increases.
Non-rotating $\ell = (0,...,7)$, $n= (1,...,n_{max})$ p-modes have been followed by progressively increasing
the rotation. This mode tracking requires special care when an avoided crossing occurs
between two modes of the same
equatorial parity. The figure on the left shows an overview of the frequency evolution while the two right figures
display zooms to illustrate avoided crossings between the $\ell =1, \; n=6$ and $\ell =5, \; n=5$ modes
and the $\ell =0, \; n=4$ and $\ell =4, \; n=3$ modes, respectively.
Although the two "interacting" modes have a mixed character near the closest frequency approach,
their original
properties are recovered after the crossing which enables to unambiguously follow and label the modes.
This is illustrated in Fig.~\ref{spec.l0v4} by considering the spectra of Legendre expansion components of the
$\ell =0, \; n=4$ and $\ell =4, \; n=3$ modes at the rotation rates marked by an arrow. Note that in the above figures
crossings do occur between equatorially symmetric and anti-symmetric modes. In the global view, there are two 
examples of discontinuous
frequency changes due to avoided crossing with modes which frequency is not represented on the figure. Actually, the 
$\ell = 8$, $n= (1,2,3)$ modes have been displayed in this view to avoid more discontinuous changes.
}
\label{tout}
\end{figure*}

\section{Results}

The parameter range of the calculations is first presented.
Then, we describe the method used to label the eigenmodes,
the structure of the frequency spectrum, some
properties of the eigenfunctions and
the differences with perturbative methods.

\subsection{Parameter range}

Self-gravitating uniformly rotating polytropes of index $N=3$ and specific heat
ratio ${\Gamma}_{1,0} = 5/3$ have been computed for rotation rates varying
from $\Omega = 0$ up to $\Omega / \Omega_K = 0.59 $.
In this range, the
flatness of the star's surface $\epsilon= 1 -R_p/R_e$ increases from $0$ to $0.15$.

Low frequency axisymmetric p-modes have been computed for each polytropic model.
We started with the non-rotating model and computed the $\ell = (0,...,7), \; n= (1,...,n_{max})$ axisymmetric p-modes,
the largest radial order depending on the degree $\ell$:
$n_{max}=10$ for $\ell=(0,1)$, $n_{max}=9$ for $\ell=(2,3,4)$
and $n_{max}=8$ for $\ell=(5,6,7)$.
All these $71$ modes were then calculated at higher rotation rates by progressively increasing
the rotation of the polytropic model.
In the next section, we explain 
how we could track and label them from zero rotation to $\Omega / \Omega_K = 0.59 $.

\begin{figure*}[htb]
\includegraphics{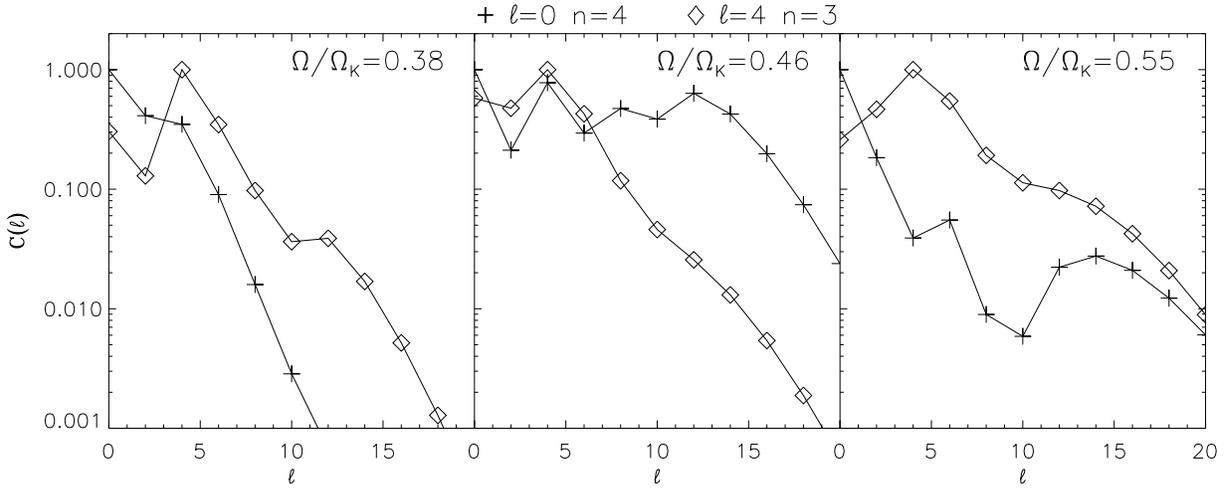}
\caption{
Evolution of the Legendre components $C(\ell) = max_{n_r} |U(\ell,n_r )|$ of the $\ell =0, \; n=4$ and $\ell =4, \; n=3$ modes
during the avoided crossing shown in Fig.~\ref{tout}. Prior to ($\Omega / \Omega_K = 0.38$) and after ($\Omega / \Omega_K = 0.55$)
the avoided crossing, the spectrum of Legendre components peaks at a given degree while, near the closest approach
($\Omega / \Omega_K = 0.46$), the double peaks
of the spectra
show the mixed character of the eigenmodes.}
\label{spec.l0v4}
\end{figure*}

\subsection{Mode labeling}

In the absence of rotation, modes are identified and classified by the three "quantum" numbers
$n, \ell, m$ characterizing  their radial, latitudinal and azimuthal structure respectively.
Because of separability, independent 1D eigenvalue problems are solved for 
each couple ($\ell$, $m$)
and it is then an easy task to order the computed frequencies, the order $n$ additionally indicating the number of
radial nodes of the mode.
By contrast, in the presence of rotation, independent 2D eigenvalue problems are solved 
for a given $\mid \! m \! \mid$ and a given equatorial parity. The computed modes are then obtained
without a priori information
about their latitudinal and radial structures.
Therefore, an important issue is whether it is possible to define a meaningful classification of these modes.
In this present work, we investigate the possibility to associate unambiguously each mode with a non-rotating mode
thus identifying it by the three quantum numbers $n, \ell, m$
of the non-rotating mode.
Similarly, \citet{Cl86} followed some equatorially symmetric acoustic modes to high rotation rates but
in a limited frequency range
and using low spatial resolution calculations.

In practice, instead of backtracking modes towards zero rotation, we
started at zero rotation with a mode we are interested in and
tried to follow it by progressively increasing
the rotation.
We did manage to track all the 
$\ell = (0,...,7)$, $n= (1,...,n_{max})$ axisymmetric p-modes from $\Omega=0$ to $\Omega / \Omega_K = 0.59$,
a global view of the eigenfrequencies evolution being displayed in Fig.~\ref{tout} (left panel).
As explained below, the main difficulty comes from avoided crossings between modes of the same equatorial parity.

Zooms in the $\omega - \Omega$ plane displayed in Fig.~\ref{tout} (right panels) provides 
two examples of avoided crossings respectively between odd ($\ell =1, \; n=6$ and
$\ell =5, \; n=5$) and even ($\ell =0, \; n=4$ and
$\ell =4, \; n=3$) modes.
Modes tends to cross because their frequencies are not affected in the same way by the centrifugal force
but, as two eigenstates of the same parity cannot be degenerate, an avoided crossing
takes place during which the
two eigenfunctions exchange their characteristics.
This exchange of property is illustrated in Fig.~\ref{spec.l0v4}
in the case of the ($\ell =0, \; n=4$),
($\ell =4, \; n=3$) crossing. 
A mean Legendre spectrum is displayed
before, near the closest frequency separation and after the avoided crossing. 
The mean Legendre spectrum of a field $U$ 
is defined as $C (\ell) = max_{n_r} |U(\ell,n_r )|/ max |U(\ell,n_r )|$, where
$U(\ell,n_r )$ are the components of the Legendre/Chebyshev expansion, $n_r$ being the degree of the Chebyshev
polynomial. The quantity $C(\ell)$ thus represents the 
largest Chebyshev component for a given value of $\ell$ normalized by the maximum over all spectral
components.
The mean Legendre spectra peak at one characteristic degree before and after
the avoided crossing, thus showing that the modes recover their original
properties after the crossing and therefore can be unambiguously recognized.
Up to the fastest rotation considered, the $\ell =0-7, \; n=1-10, \; m=0$, p-modes
undergo a limited number of avoided crossing and could be followed
unambiguously.

It remains that near the crossing the labeling
is somewhat ambiguous. First, it is difficult to define
a criterion to assign a label. Here, we mostly use the degree at which the mean Legendre spectrum reaches a maximum.
But it occurred that the two interacting modes peaks at the same degree
in which case we determined the location of the smallest frequency separation.
Second, as shown by Fig.~\ref{spec.l0v4}, a more fundamental problem is
that a single label can not reflect the mixed nature of the eigenfunction.

\begin{figure*}
\includegraphics{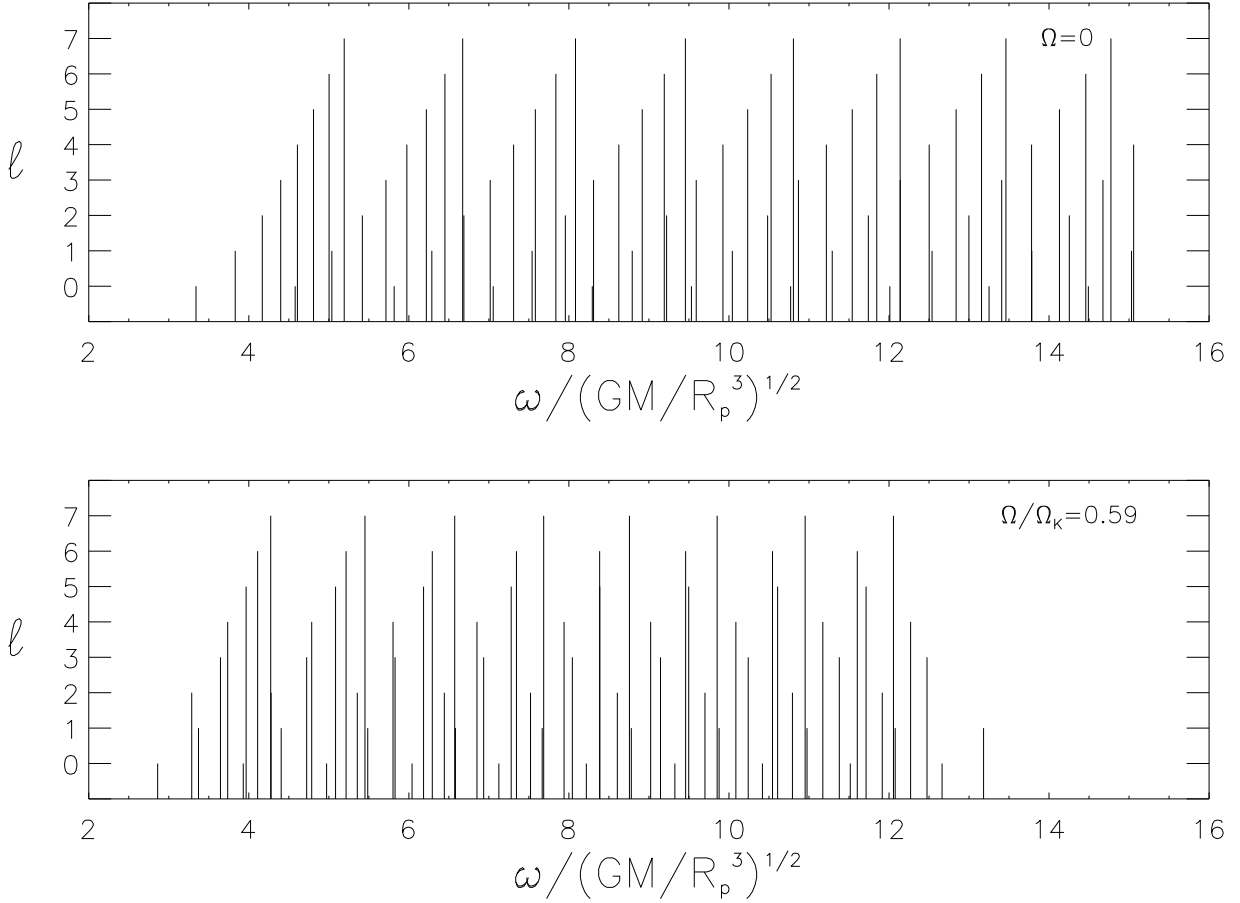}
\caption{
Frequency spectrum of $\ell =0-7, \; n=1-n_{max}, m=0$ modes at $\Omega = 0$ (top panel) and
$\Omega / \Omega_K =0.59$ (bottom panel). The degree number $\ell$ associated with the frequency is shown by the height of
 the vertical bar.}
\label{spectrum}
\end{figure*}

Another issue related to avoided crossings concerns
their influence on the 
accuracy of the eigenfunction computation. Indeed, if a large-scale well
resolved eigenfunction undergoes an avoided crossing with
a small scale unresolved mode,
the accuracy of the eigenfunction determination will be affected.
The effect on the frequency accuracy should be small 
as the frequency gap induced by the 
avoided crossing of two modes of well separated length scales
is small.
But, right at the closest approach, the eigenfunctions
will be much affected.
At zero rotation, the highest degree mode present in our frequency range
is $\ell = 51 , n=1$. Thus, if one of the low degree mode that we computed undergoes an avoided crossing with a mode
of such a high degree, the high degree mode should be resolved to ensure an accurate determination of the
eigenfunction of the low degree mode.

\subsection{The structure of the frequency spectrum}

The effect of the centrifugal force on the acoustic frequency spectrum of axisymmetric modes
is investigated. The mean modifications of the spectrum are first commented.
Then, we investigate how regularities in the frequency spacings
evolve with rotation.  Finally, differences between equatorially symmetric
and anti-symmetric modes are outlined.

\subsubsection{Global spectrum evolution}

Figure~\ref{spectrum} compares the frequency spectrum of the $\ell =0-7, n=1-n_{max}, m=0$ modes at $\Omega = 0$
(upper panel) and at $\Omega / \Omega_K =0.59$ (lower panel), the height of the vertical bars corresponding
to the degree $\ell$ of the mode.
It appears that the centrifugal force induces a mean contraction of the frequency spectrum.
This is expected as the decrease of the sound speed and the increase of the star volume
induced by the centrifugal force both tend to lessen the frequency of acoustic modes.

To illustrate this effect, one could think of a spherically symmetric decrease of the effective gravity.
If one considers a homologous series of spherical models of increased volume $V$, the decreasing rate of the frequencies
$\Delta \omega/\omega$ is $-(1/2) (\Delta V/V)$, as the normalized frequencies 
$\omega/(GM/R^3)^{1/2}$ remain constant.
For non-homologous spherically symmetric changes, $\Delta \omega/\omega$ is asymptotically equal to $- \Delta (\ln \int_0^{R_p} dr/c)$
for high order modes verifying the following asymptotic formula valid for low degree and high order p-modes \citep{Ta80}:
\begin{equation}
\omega = \frac{\pi}{\int_0^R \frac{dr}{c}} (n + (\ell + 1/2)/2 + \alpha)
\label{eq:asym}
\end{equation}
\noi
where $1/\int_0^R \frac{dr}{c}$ is the sound travel time along a stellar radius and $\alpha$ is a constant.
When, as in these two previous cases, $\Delta \omega/\omega$ does not depend on the frequency, the concentration of the frequency spectrum 
is homothetic.

\begin{figure*}
\resizebox{\hsize}{!}{\includegraphics{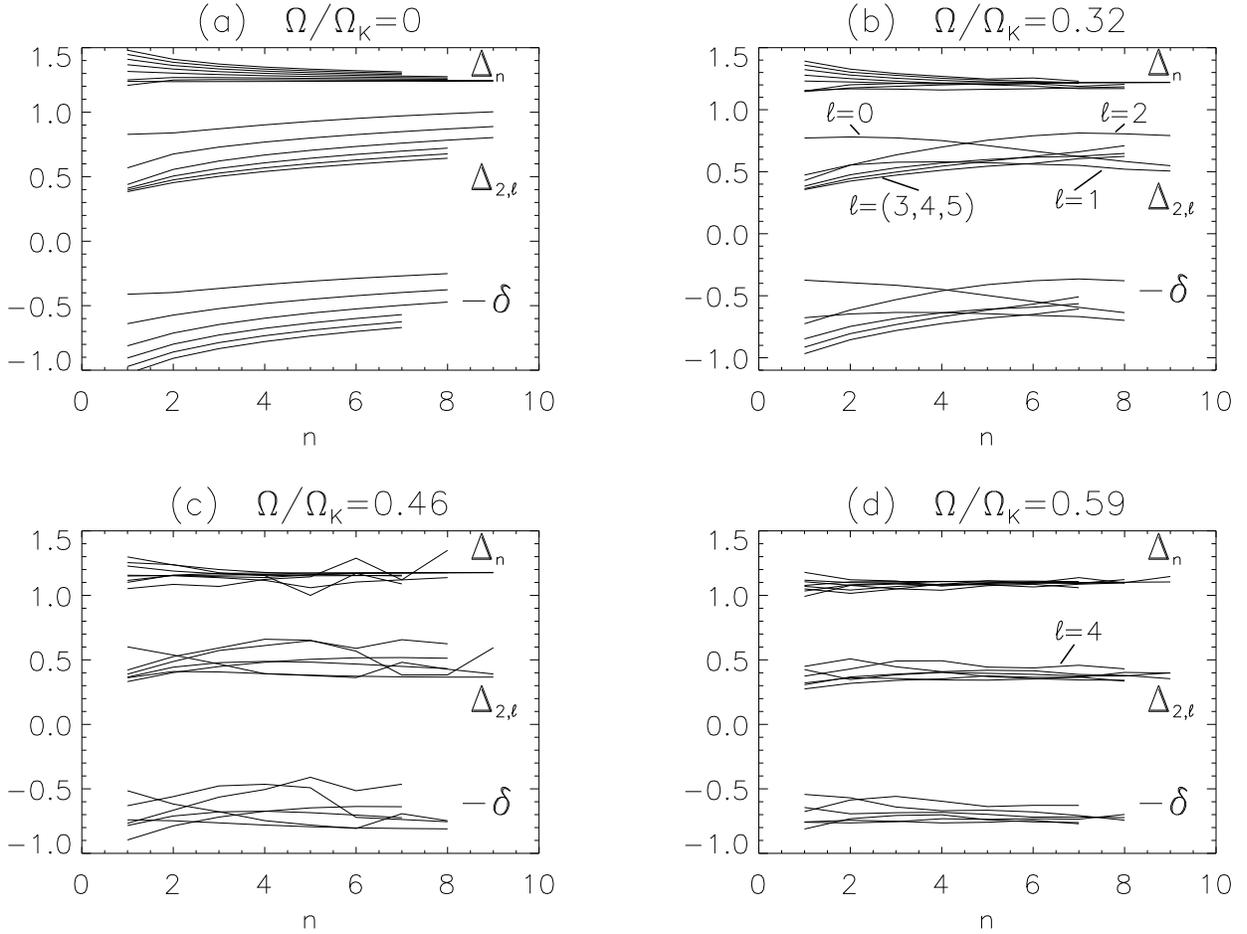}}
\caption{Regularities in the frequency spacings of axisymmetric ($m=0$) modes.
The large frequency separation between modes of consecutive order $\Delta_n = \omega_{n \ell} - \omega_{n-1 \ell}$,
the frequency separation between $\ell+2$ and $\ell$ modes,
$\Delta_{2,\ell}= \omega_{n,\ell+2} -  \omega_{n,\ell}$, and the small frequency separation $\delta = \Delta_n - \Delta_{2,\ell}$ 
are displayed as a function of the radial order $n$
for
four different rotation rates, (a) $\Omega = 0$, (b) $\Omega / \Omega_K =0.32$,
(c) $\Omega / \Omega_K =0.46$ and
(d) $\Omega / \Omega_K =0.59$. We plotted the opposite of the small frequency separation $-\delta$ for the clarity of the figure.
Continuous lines have been drawn between frequencies of the same
degree $\ell$.}
\label{diff}
\end{figure*}

This is clearly not the case here since the frequencies cross each other (see Fig.~\ref{tout}).
But there is still an average contraction rate which is of the order of $-(1/2) (\Delta V/V)$, where now $V$ is the volume
of the centrifugally distorted star.
In addition, the contraction rates of individual frequencies appears to be comprised
between the logarithmic derivative of the sound travel times computed respectively
along the polar and equatorial radii:
\begin{equation}
\partial_{\Omega} \left(\ln \int_0^{R_p} \frac{dr}{c} \right) \leq - \partial_{\Omega} (\ln \omega) \leq
\partial_{\Omega} \left(\ln \int_0^{R_e} \frac{dr}{c}\right)
\end{equation}
Another interesting property is that, at small rotation rates say $\Omega/\Omega_{K} \leq 0.05$,
the contraction rate $\partial_{\Omega} (\ln \omega)$ tends to be independent of
$\ell$ and $n$ for the large degree modes $\ell \geq 3$.
This suggests that an asymptotic regime exists for modes with horizontal wavelengths
smaller than the dominant length scales of the
centrifugal distortion. In this regime, the contraction rate has a constant value which is not equal to $-(1/2) (\Delta V/V)$
and that would be interesting to determine.
We already found such behaviour in the case of homogeneous ellipsoids \citep{Li01} where
a
perturbative analysis shows that the contraction rate of axisymmetric modes is constant for high $\ell$ and $n$ and that it can be
related to the increase of the ellipse perimeter.

Nevertheless, for the low degree modes $\ell \leq 2$ below $\Omega/\Omega_{K} \approx 0.05$ and for all modes at higher rotation rates,
$\partial_{\Omega} (\ln \omega)$ depends on $\ell$ and $n$. This
differential effect modifies the structure of the frequency spectrum as the rotation increases.

\subsubsection{Regular frequency spacings}

In a non-rotating star, the frequency spectrum presents some regular frequency spacings which can be accounted
for by an asymptotic theory in the high frequency limit $\omega \rightarrow \infty$.
The asymptotic formula \eq{eq:asym}, valid for low degree and high order modes,
shows that 
the large frequency
separation between modes of consecutive order $n$ , $\Delta_n = \omega_{n+1, \ell} - \omega_{n, \ell}$,
does not depend on $\ell$ and $n$ and is equal to $\pi / \int_0^R \frac{dr}{c}$.
A more detailed asymptotic analysis also shows how the 
so-called small frequency separation $\delta = 
\omega_{n+1, \ell} - \omega_{n, \ell+2}$ vanishes as a function of the 
frequency.
Although our calculations are restricted to the low frequency part of the acoustic spectrum, 
we do observe a clear tendency towards these
asymptotic behaviors in the non-rotating case.
We can therefore investigate whether
these properties are modified by rotation.

Figure~\ref{diff} presents the large frequency separation $\Delta_{n}$
and the frequency separation between consecutive modes of the same order and parity:
\begin{equation}
\Delta_{2,\ell}= \omega_{n,\ell+2} -  \omega_{n,\ell},
\end{equation}
\noindent as a function of the radial order $n$
for
four different rotation rates, (a) $\Omega = 0$, (b) $\Omega / \Omega_K =0.32$,
(c) $\Omega / \Omega_K =0.46$ and
(d) $\Omega / \Omega_K =0.59$.
As in the previous figures, the frequencies are adimensionalized by $(GM/R_p^3)^{1/2}$. Continuous lines have 
been drawn between frequencies of the same
degree $\ell$.
We first observe that the large frequency separation tends to be independent of $n$ and $\ell$
at all rotation rates.
In accordance with the mean contraction of the
frequency spectrum mentioned above, the large frequency separation
decreases with rotation. It is always comprised between $\pi / \int_0^{R_p} dr/c$ and
$\pi / \int_0^{R_e} dr/c$.

The dispersion of the large frequency separations around their mean value also has an interesting evolution with rotation.
In the non-rotating case, the dispersion reflects a regular departure from the asymptotic limit.
It is larger for high degrees and monotonically decreases
with frequency (see Fig.~\ref{diff}a).
In the rotating cases, the dispersion is not as 
regular.
The largest departures, some of which are most clearly visible in Fig.~\ref{diff}c, 
can be attributed to an ongoing avoided crossing.
The residual dispersion is irregular and decreases with rotation.
At $\Omega / \Omega_K =0.59$, if we exclude all $n < 4$ values from our sample, the 
mean large frequency separation $<\Delta_n>$ is equal to $1.095 (GM/R_p^3)^{1/2}$
and its standard deviation is $0.017 <\Delta_n>$.

We now consider the small frequency separation
$\delta = \Delta_n - \Delta_{2,\ell}$.
As expected, in the absence
of rotation the small frequency separation tends to vanish as $n$ increases.
But, at $\Omega / \Omega_K =0.32$, 
the small frequency separation no longer decreases with $n$ for some values of $\ell$
and for the higher rotation rates it becomes nearly constant.
At the same time, 
the $\Delta_{2,\ell}$ separation becomes more and more uniform
as rotation increases. As shown in Fig.~\ref{diff}b, $\Delta_{2,\ell}$ becomes approximatively constant with $n$ first for low degree modes
while it still increases with frequency
for high degree modes. In addition, equatorially antisymmetric modes reach
this new regime at a lower
rotation rate than the equatorially symmetric modes of similar degree.
This is illustrated in Fig.~\ref{diff}d by the $\Delta_{2,\ell=4}$ curve
which 
still remains above the mean $\Delta_{2,\ell}$ value while the $\Delta_{2,\ell=5}$
separation already collapses with the other curves.
At $\Omega / \Omega_K =0.59$, if we exclude all $n < 4$ values from our sample, 
the mean frequency separation $\Delta_{2,\ell}$ is equal to $0.387 (GM/R_p^3)^{1/2}$ and its
standard deviation is
$0.033 <\Delta_n>$.

As a consequence of the near uniformity of $\Delta_n$ and $\Delta_{2,\ell}$,
the frequencies of low degree and high order modes can be approximated
by the following expressions:
\begin{equation}\label{eq:latate}
\tilde{\omega}_{n \ell} = \left\{\begin{array}{l}
n \delta_n + p \delta_{\ell} + \alpha^+ \;\mbox{if} \; \ell = 2p \\
n \delta_n + p \delta_{\ell} + \alpha^- \;\mbox{if} \; \ell = 2p + 1 \end{array}\right.
\end{equation}
\noi
where $\delta_n=<\Delta_n>$, $\delta_{\ell}=<\Delta_{2,\ell}>$, 
$\alpha^+$ and $\alpha^-$ only depend on the equilibrium model.
Using a reference frequency to determine the $\alpha$ constants (the $\ell=0, \; n=8$ frequency
for $\alpha^+$ and the $\ell=1, \; n=8$ frequency for $\alpha^-$), we computed the root mean square error
$\sqrt{1/N \sum ({\tilde{\omega}} - \omega)^2}$ and
the maximum error made in using the approximate expressions \eq{eq:latate}. 
For a frequency subset containing the $n>4$ and $\ell<5$ modes, the rms error is 
$0.017 \delta_n$ while the maximum error amounts to $0.05 \delta_n$. Both errors are a very small fraction of 
the large separation which shows that Eq. \eq{eq:latate} yields
useful approximations of the frequency spectrum.

\subsubsection{Equatorially symmetric versus anti-symmetric frequency spectra}

We have seen that the regular frequency spacings $\Delta_n$ and $\Delta_{2,\ell}$ have similar
values for symmetric and anti-symmetric modes with respect to the equator.
The evolution of the equatorially symmetric and anti-symmetric frequency spectra are nevertheless
quite different. Indeed, considering two modes of similar frequency but of opposite equatorial parity,
the frequency of the symmetric mode 
generally decreases faster with rotation than the frequency of the antisymmetric modes.
The consequence is that the frequency separation between modes of consecutive degree (and thus
of opposite parity)
$\Delta_l = \omega_{n,\ell+1} -  \omega_{n,\ell}$
tends to increase when $\ell$ is even and to decrease when $\ell$ is odd.
The frequency separation $\Delta_l$ can even become negative which
implies that, contrary to the non-rotating case, frequencies of a given order $n$ do not increase
monotonically with the degree $\ell$.
This striking modification of the usual frequency ordering is apparent on
Fig.~\ref{spectrum} where the $(\ell =2 , n)$ frequencies are smaller than the $(\ell =1 , n)$ frequencies
for all the order $n$ that we calculated, that is $n= (1,...,10)$. In the same way, the $(\ell =4 , n)$ frequencies are smaller than the
$(\ell =3 , n)$ frequencies if $n \geq 3$, and again the $(\ell =6 , n)$ frequencies 
are smaller than the $(\ell =5 , n)$ frequencies if $n \geq 5$.

\subsection{Equatorial concentration}

In this section, 
we shall focus on the most notable effect of the centrifugal force on the eigenmodes,
namely the equatorial concentration and consider its consequences on the mode visibility. 
Note that \citet{Cl81} also reported an equatorial concentration of
the equatorially symmetric modes that he calculated.

\begin{figure*}
\resizebox{\hsize}{!}{\includegraphics{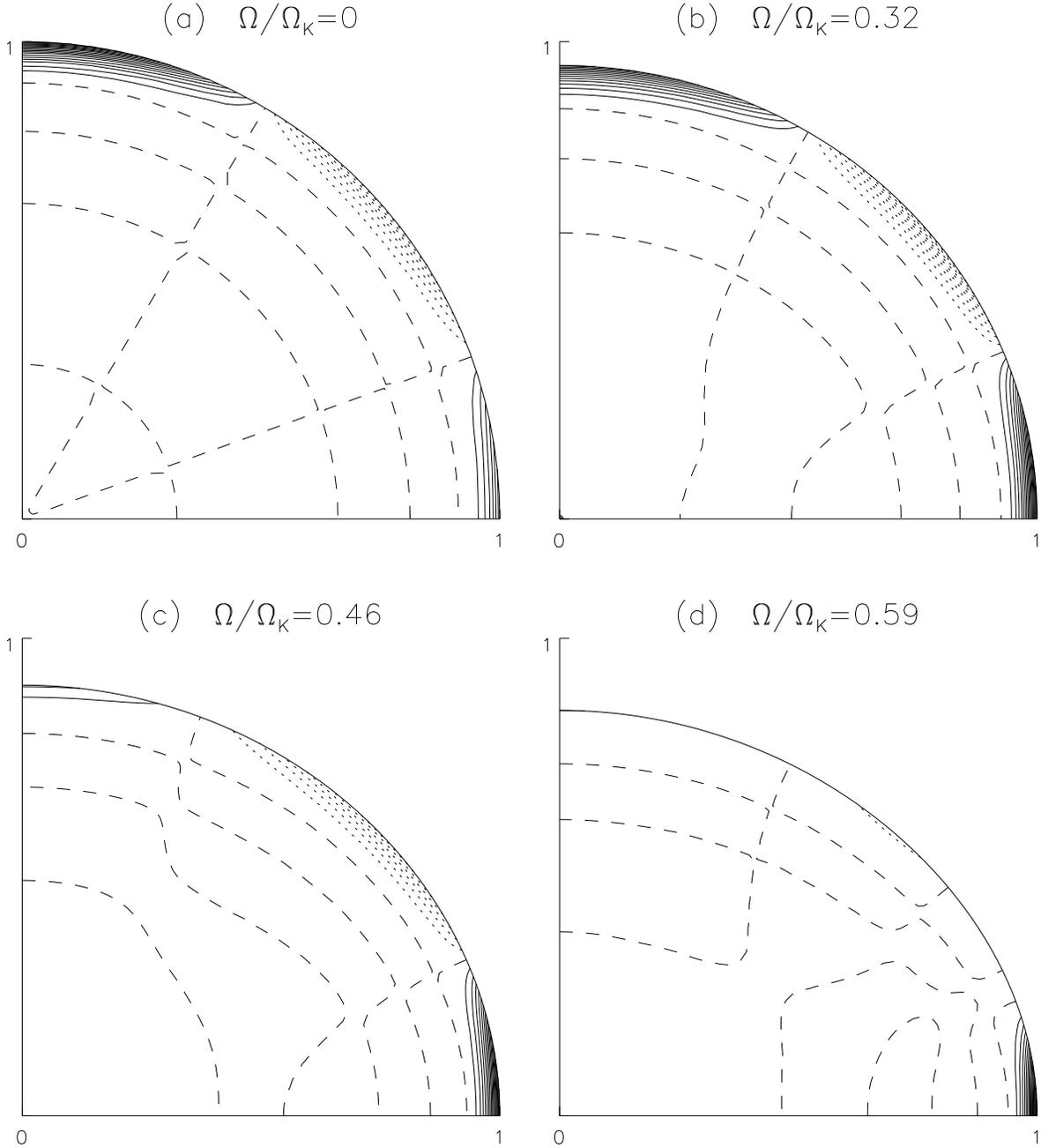}}
\caption{Isocontours of the $\ell=4, n=4$ mode amplitude in a meridional plane as a function of the rotation
(a) $\Omega = 0$, (b) $\Omega / \Omega_K =0.32$,
(c) $\Omega / \Omega_K =0.46$ and
(d) $\Omega / \Omega_K =0.59$. The amplitude is normalized with the maximum of its absolute value.
Continuous lines correspond to positive amplitudes, dashed lines to the zero amplitude and dotted lines
to negative amplitudes. At zero rotation, the angular distribution is given by the
$\hat{Y}^0_4 (\theta)$ Legendre polynomial while the radial distribution is characterized
by the surface concentration of the amplitude and the presence of $n=4$ nodes in the inner part. 
For larger rotation rates, the largest amplitudes concentrates toward the equator. We also
note that the number of radial nodes decreases along the polar radius while it increases along the equatorial
radius.}
\label{modes}
\end{figure*}

Figure~\ref{modes} 
shows this effect on the ($\ell=4, n=4$) mode. Contours of the amplitude of the Lagrangian pressure perturbation 
are plotted in a meridional plane
for increased rotation rates, (a) $\Omega = 0$, (b) $\Omega / \Omega_K =0.32$,
(c) $\Omega / \Omega_K  =0.46$ and
(d) $\Omega / \Omega_K  =0.59$.
We observe that the number of nodes increases along the equatorial radius and decreases along the polar one.
Along the surface, the number of nodes remains equal to $\ell$ before $\Omega / \Omega_K  =0.59$
where additional nodes appear.
The equatorial concentration is clearly seen in the outermost layers.

\begin{figure*}
\resizebox{\hsize}{!}{\includegraphics{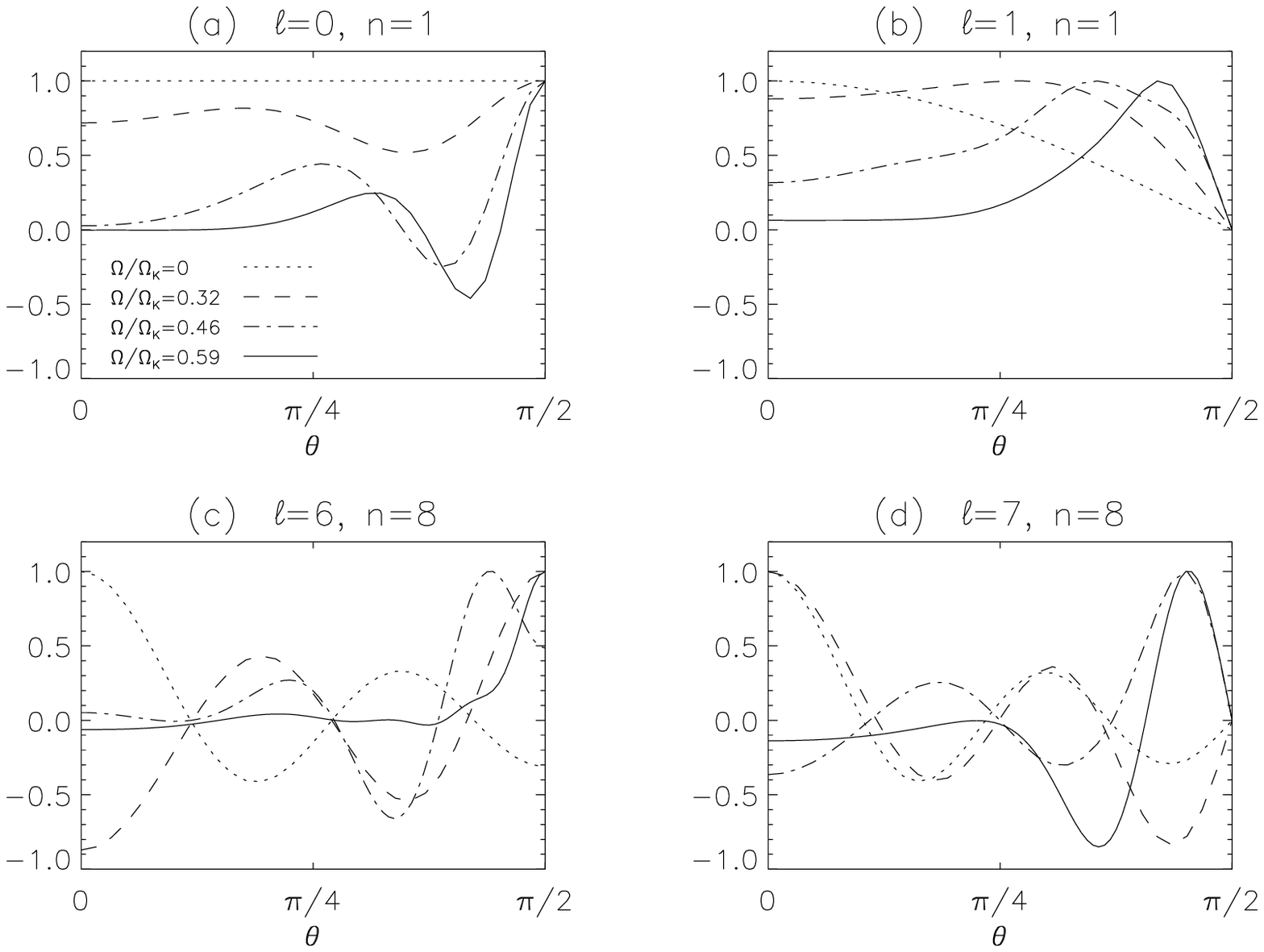}}
\caption{The mode amplitude at the surface of the polytropic model as a function of the rotation rate (a) $\ell=0, n=1$, (b) $\ell=1, n=1$,
(c) $\ell=6, n=8$, (d) $\ell=7, n=8$. The amplitude is normalized with the maximum of its absolute value.
While the angular distribution is given by the corresponding Legendre polynomial $\hat{Y}^0_\ell (\theta)$ in the absence
of rotation, the oscillation amplitude progressively concentrates towards the equator ($\theta = \pi/2$)
as rotation increases.}
\label{conc}
\end{figure*}

In Fig.~\ref{conc}, the equatorial concentration is shown for other modes including
the lowest and highest degree modes of our sample as well as
symmetric and anti-symmetric modes. The latitudinal variation of the mode amplitude
is displayed at the surface for the following modes, (a) $\ell=0, n=1$, (b) $\ell=1, n=1$,
(c) $\ell=6, n=8$, (d) $\ell=7, n=8$.
In each case, the equatorial concentration grows with rotation.
At the largest rotation rate, symmetric modes are maximal at the equator
while anti-symmetric modes peak at small latitudes since they must vanish at the equator.
The contrast between these maxima and the polar
amplitude is strong.

The equatorial concentration reveals a modification of the resonant cavity of the acoustic waves.
In particular,
the reduction of the volume of the resonant cavity
should
tend to increase the frequency.
The equatorial concentration seems also to be associated with the near-uniformity 
of the frequency separation $\Delta_{2,\ell}$.  
At small rotation rates, the concentration is not completed and $\Delta_{2,\ell}$ is clearly not constant.
Whereas, at the largest rotation rate, all modes are concentrated near the equator and $\Delta_{2,\ell}$ is nearly uniform.

\begin{figure*}
\resizebox{\hsize}{!}
{\includegraphics{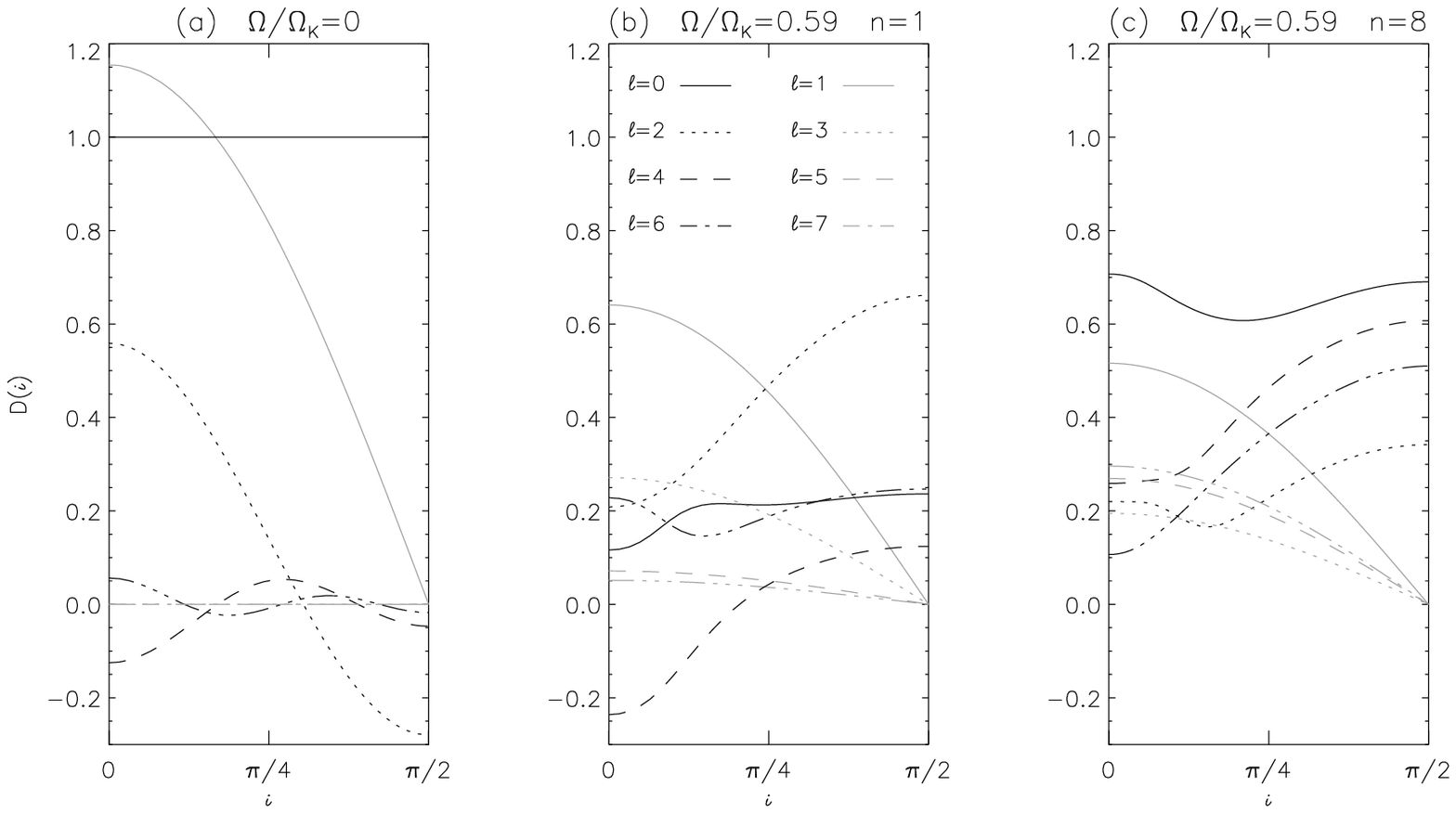}}
\caption{The disk-averaging factor $D(i)$ is shown as a function of the inclination angle $i$ for various axisymmetric
modes at two different rotation rates $\Omega = 0$ (a) and $\Omega / \Omega_K =0.59$.
The degree of the modes varies from $\ell=0$ to $\ell=7$. In the rotating case, the surface distribution also depends on the
order of the mode. Two values $n=1$ (b) and $n=8$ (c) have been considered at $\Omega / \Omega_K =0.59$.}
\label{inclin}
\end{figure*}

\begin{figure*}
\resizebox{\hsize}{!}
{
\includegraphics{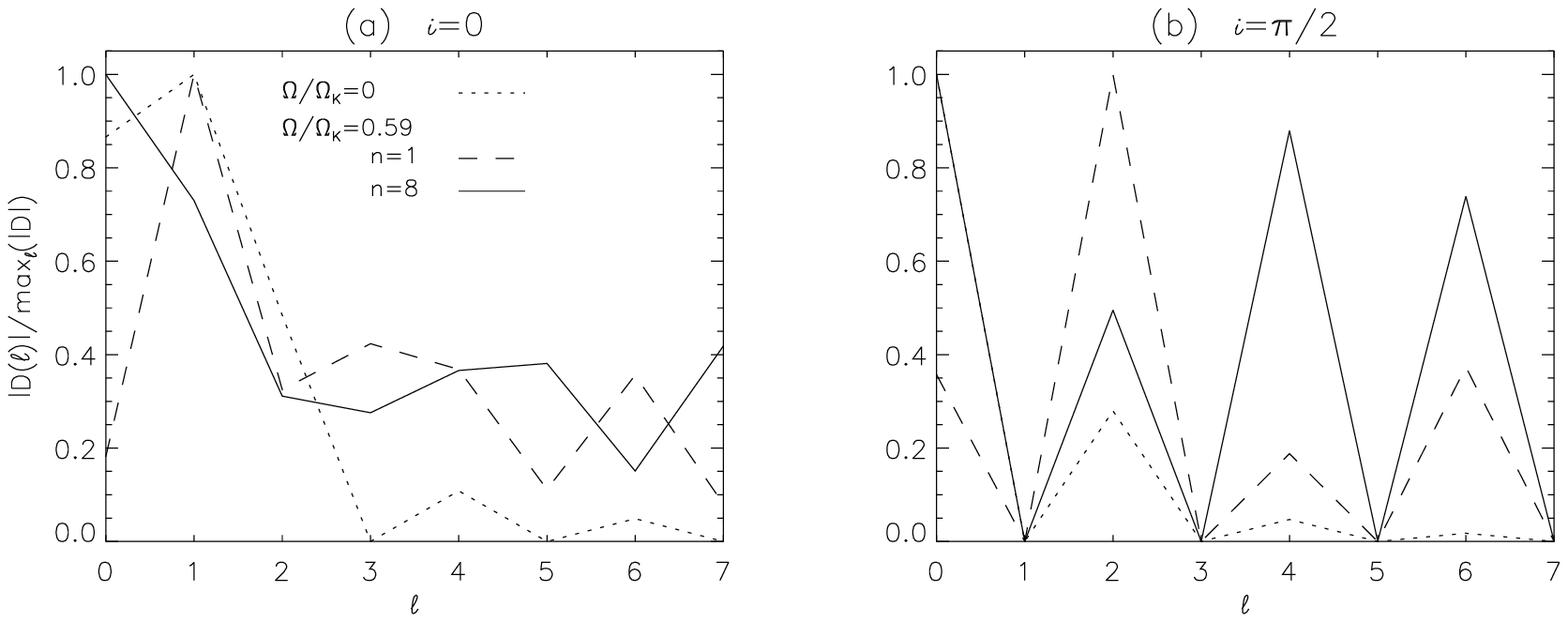}}
\caption{The variation of the disk-averaging factor as a function of the degree $\ell$ is shown at two fixed
values of the inclination angle ($i = 0$ and $i= \pi/2$). The three curves
on each figure
correspond to the non-rotating case and, at $\Omega / \Omega_K =0.59$, to
two different orders, $n=1$ and $n=8$.
The absolute value of disk-averaging factor has been rescaled by its maximum value among the different
degree considered to outline its dependency on the degree number.
In sharp contrast with the non-rotating case, the disk-averaging factor at $\Omega / \Omega_K =0.59$ does not show a strong
decrease with $\ell$.}
\label{l_ratio}
\end{figure*}

Besides its effect on the frequency spectrum, the equatorial concentration of eigenmodes should induce
a profound modification of the mode visibility as compared to the non-rotating case.
The photometric mode visibility is determined by the integration over the visible part of the star's perturbed surface
of the radiation intensity perturbations associated with a particular pulsation mode.
Rigorous calculations of photometric visibilities are beyond the scope of the present paper as
they require non-adiabatic calculations of the oscillation modes and stellar atmosphere models (e.g. Daszy\'nska-Daszkiewicz et al. 2002 ).
But we can still determine the effects of averaging the perturbations over the visible surface which have a direct impact
on the visibility. The disk-averaging factor is defined as:
\begin{equation}\label{eq:disk}
D(i)= 
\frac{1}{\pi R_e^2 \delta T_0} \int\!\!\!\!\int_{S_v} \delta T (\theta,\phi) {\bf d S} \cdot {\bf e_i}
\end{equation}
\noi
where $i$ is the inclination angle between the line-of-sight and the rotation axis, ${\bf e_i}$ is a unit vector in the 
observer's direction and $\delta T$ is the spatial part of the Lagrangian temperature perturbation 
at the stellar surface, $\delta T$ being proportional to the velocity divergence $\chi$ 
in the approximation
of adiabatic perturbations.  The mode amplitude is normalized by $\delta T_0$ the root mean square of the perturbation over the whole stellar surface
\begin{equation}\label{eq:t0}
\delta T_0 = \lp \int\!\!\!\!\int_{S} \delta T^2 (\theta,\phi) d S \rp^{1/2}
\end{equation}
\noi
and the visible surface $S_v$ has been normalized by $\pi R_e^2$, the visible surface of a star seen pole-on.
With these normalizations the disk-averaging factor of a radial mode seen pole-on is unity.

In the absence of rotation, the surface distribution of modes is determined by a unique spherical harmonic and the disk-averaging factor takes a simple
analytical form \citep{Dz77}. For even degree and for $\ell =1$, the disk-averaging factor varies with
the inclination 
angle as the Legendre polynomial $\YTL(i)$
while it vanishes altogether for odd degree $\ell \ge 3$.
For rotating stars, the method of the calculation is detailed in Appendix C. 
Note that for modes which are equatorially anti-symmetric and axisymmetric,
the disk-averaging factor has also a simple dependency on the 
inclination angle
as it is proportional to $\cos(i)$.

Figure~\ref{inclin} shows the disk-averaging factor of various axisymmetric modes as a function of the inclination angle.
The non-rotating case is displayed in Fig.~\ref{inclin}a where $\ell =0$ to $\ell=7$ modes are considered. We recall that,
at $\Omega = 0$, 
modes of different radial orders but same $\ell$ and $m$ have the same surface distribution.
This is not true in the rotating case and, at $\Omega / \Omega_K =0.59$, Fig.~\ref{inclin}b and \ref{inclin}c 
present the disk-averaging factor
for modes of the same degree numbers but for two different radial orders $n=1$ and $n=8$, respectively.
Note also that the disk-averaging factor was allowed to take negative value for the clarity of the figure although
it is its absolute value which is relevant for the mode's visibility.
Figure~\ref{inclin} shows that rotation strongly modifies the dependency of the disk-averaging factor on the inclination angle as well as
on the degree number.

Fig.~\ref{inclin}c shows that for all $n=8$ equatorially symmetric modes 
the absolute value of disk-averaging factor tends to increase 
with the inclination angle. This is due to 
the equatorial concentration of these modes (see for example the surface
distribution of the $(\ell=6, n=8)$ mode at $\Omega / \Omega_K =0.59$ shown in Fig.~\ref{conc}c). 
This tendency is less
pronounced for the $n=1$ symmetric modes shown in Fig.~\ref{inclin}b (except for the $\ell=2$ mode) although these modes
are also equatorially concentrated. This is due to a cancellation effect between positive and negative perturbations
concentrated near the equator as illustrated by the surface
distribution of the $(\ell=0, n=1)$ mode in Fig.~\ref{conc}a.
The non-rotating case strongly differs since the absolute value of the disk-averaging factor for even degree $\ell > 0$ modes does not vary
monotonically with the inclination. Indeed, they have $\ell/2$ nodes between $0$ and $\pi/2$.
For odd $\ell$ modes, the disk-averaging factor is also modified by rotation since it no longer vanishes for $\ell \ge 3$.
This occurs because the projected elementary surfaces ${\bf d S} \cdot {\bf e_i}$ are no longer symmetric with respect to the 
observer's direction and because the projection of the eigenmode surface distribution onto the Legendre polynomial 
${\hat{Y}^0_1}$ 
is not zero for $\ell \ge 3$ modes.

In non-rotating stars, the cancellation effect between positive and negative perturbations results in a rapid
decrease of the disk-averaging factor as the degree $\ell$ of the mode increases. Consequently, modes above a certain degree $\ell \geq 3-4$ are not 
expected to be detectable with photometry and are therefore not included when trying to identify the observed frequencies.
As shown in Fig.~\ref{l_ratio}, this property must be reconsidered 
for rapidly
rotating stars. The absolute disk-averaging factor normalized
by its maximum value over the degree considered $0 \leq \ell \leq 7$, $|D(i)|/max_{\ell}|D(i)|$,
is plotted as a function of $\ell$ for two fixed values of the inclination angle, $i = 0$ on Fig.~\ref{l_ratio}a and $i= \pi/2$
on Fig.~\ref{l_ratio}b. The three curves correspond to $\Omega =0$ and to $\Omega / \Omega_K =0.59$ 
for the $n=1$ and $n=8$ modes, respectively.
In sharp contrast with the non-rotating case, the disk-averaging factor has no tendency to decrease above $\ell = 2$.
Again, this can be explained by the equatorial concentration as modes of different degree have a similar
surface distribution.

\subsection{Comparison with perturbative methods}

\begin{figure}[htb]
\resizebox{\hsize}{!}
{
\includegraphics{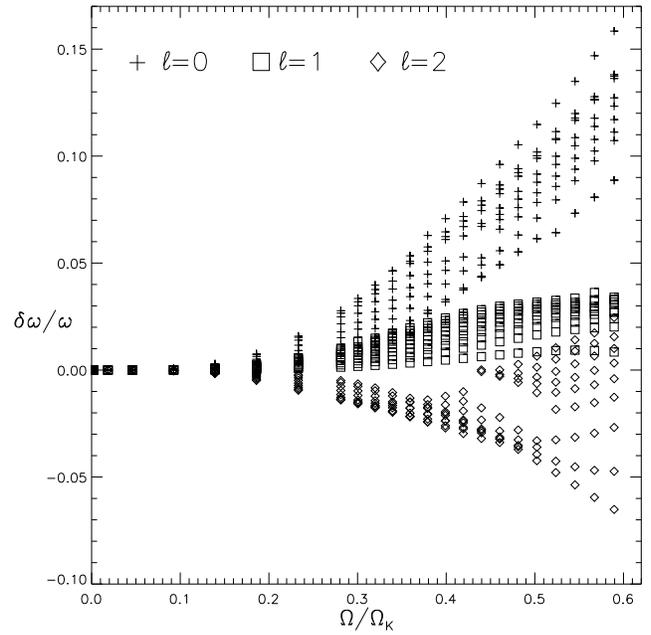}
}
\caption{
The relative difference between exact frequencies and their second order perturbative approximation
(second order in terms of the small parameter $\Omega / \Omega_K$) namely,
$\delta \omega / \omega$  where $\delta \omega = \omega - \omega_{\mbox{pert}}$
is displayed as a function of the rotation rate for the $\ell =0-2, \; n=1-10, \; m=0$
modes.}
\label{devlim}
\end{figure}

According to the perturbative analysis, centrifugal effects appear at second order in
$\Omega$ \citep{Sa81}. To determine the second-order perturbative coefficient from our complete
calculations, we performed a serie of calculations for small rotation rates ($\Omega= 0,1.8 \times 10^{-3}, 1.8 \times 10^{-2}, 4.6 \times 10^{-2},
0.09 \times 10^{-2}$, ... in units of $\Omega_{K}$). 
From them, we determined the second-order perturbative coefficient, denoted $\omega_1$, as 
the limit of the ratio $(\omega(\Omega) - \omega_0)/ \Omega^2$, where $\omega_0$ denotes a non-rotating eigenfrequency.
Thus the approximate frequencies valid up to the second order in $\Omega$ read
$\omega_{\mbox{pert}} = \omega_0 + \omega_1 \Omega^2$, where the frequencies are in units of $(GM/R_p^3)^{1/2}$
and the rotation is in unit of
$\Omega_{K}$.
To
assess the range of validity
of the second order perturbative approach,
we compared these approximate frequencies to the "exact" frequencies.
In Fig.~\ref{devlim}, the relative differences between the two calculations, $(\omega - \omega_{\mbox{pert}})/\omega$,
is plotted as a function of the rotation rate
for the $\ell =0-2, \; n=1-10$ modes.
The departures computed
for the other modes,
$\ell = (3,...,7)$, $n= (1,...,n_{max})$, are smaller than
the extremal differences shown in Fig.~\ref{devlim}
and are not displayed for the clarity of the figure.
The relative differences
are generally larger for low degree modes and, for small rotation rates, are a monotonic function
of the radial order $n$ (an increasing function for the $\ell = 0-2$ modes shown in
Fig.~\ref{devlim}).
As mentioned before the low degree modes seem to be sensitive to the precise form of the distortion
which occurs at similar lengthscales. As rotation increases, it appears that higher than second order effects are
important
to describe the effect of the centrifugal distortion on these modes.
The second order approximation is much better for large $\ell$ modes which are sensitive to global distortion properties.

As compared to the observational uncertainties
on
the frequency determinations, the error made in using second order perturbative methods becomes rapidly significant
as rotation increases.
For a ratio $\Omega/\Omega_K = 0.24$, corresponding to a typical $\delta$ Scuti star with an equatorial
velocity of $100$ \kms, the maximum
absolute difference amounts to $11 \mu \mbox{Hz}$ which is much
larger than typical observational uncertainties.

The fact that in our frequency sample the absolute difference increases with frequency
suggests that departure from the perturbative approach could be detectable for moderately rotating stars
pulsating in high order modes.
In our data limited to $n \leq 10$, the largest relative difference is $2.87 \times 10^{-3}$
for a $\Omega/\Omega_K =0.139$ rotation corresponding to a solar-type star with an equatorial
velocity of $60$ \kms.
If
we assume that the same relative difference holds for
high order p-modes in the range of $2000 \mu \mbox{Hz}$
generated in convective envelopes of these stars,
the absolute difference is $5.7 \mu \mbox{Hz}$ for a typical $2000 \mu \mbox{Hz}$ frequency.
This would be also easily detectable
given the observational uncertainties \citep{Ba05}. However, a firm conclusion should await
a direct comparison between the perturbative approach and the complete
calculation for p-modes generated in the convective envelope of rotating solar-type
stars.

\section{Discussion and conclusion}

A new non perturbative method to compute accurate oscillation modes in rotating stars has been presented.
The accuracy of the computed frequencies has been obtained by testing the effect of the different parameters
of the numerical method.
Then, the effects of the centrifugal force on
low frequency axisymmetric acoustic oscillation modes have been investigated in uniformly rotating polytropic models of stars.
Seventy-one low degree $\ell \leq 7$  and low order $n \leq 10$ modes have been first determined
at zero rotation and then tracked at higher rotation rates 
up to $\Omega/ \Omega_K = 0.59$.

In the frequency and rotation ranges considered in this paper, 
the zero rotation quantum numbers $\ell$ and $n$ have been used to label the modes.
This labeling turned out to be 
meaningful since we found regular frequency spacings between modes of the same degree
and consecutive orders, $\Delta_n$, and, within the subsets of modes of the same equatorial parity,
between modes of the same order
and consecutive degree, $\Delta_{2,\ell}$.
We noted however that near avoided crossings, when the eigenfunction is a 
mix of the two "interacting" modes, a unique label cannot reflect the actual eigenfunction
structure.
Although successful in the frequency and rotation ranges considered, it remains
to be proved that this labeling can be performed in practice at higher rotation and
at
higher frequencies.
Indeed, the main difficulty of the labeling procedure arises from the avoided crossing
between modes of the same equatorial parity and such crossings will be more frequent as
the eigenfrequency density increases with the frequency.
The coupling between modes is also stronger at higher rotation rates. 
It might then be necessary to investigate other tools than the mean Legendre power spectrum to
characterize the modes.

The study of the frequency spectrum showed a quite unexpected result,
namely that, at the highest rotation rates, a new form of organization sets in after the
zero rotation asymptotic structure of the spectrum has been destroyed.
In the absence of rotation, the asymptotic theory is directly related to
the spherical symmetry of the stars and ultimately to the integrability of
underlying ray dynamics. In the presence of rotation,
the eigenvalue problem is not fully separable and the
underlying acoustic ray dynamics is most probably not integrable.
Then, the regular spacings observed at high rotation rates were not really expected.
They might be the sign of a near-integrable
ray dynamic rather than a chaotic system.
These aspects will be investigated through a ray dynamic study of
rotating polytropic models of stars.

Most importantly for asteroseismology, the existence of regular spacings in the spectrum can potentially provide
tools for the mode identification in rapidly rotating pulsating stars.
A complete acoustic frequency spectrum including $m \ne 0$ modes and the effects
of
the Coriolis acceleration
should however
be computed and analyzed to assess the practical relevance of these
regular spacings.

Apparently, there is a relation between the new spectrum structure 
and
the equatorial concentration of the mode amplitudes.
A consequence would be that 
this spectrum structure does not apply
to the whole spectrum.
Indeed, sufficiently high degree modes should still be of the
whispering gallery type (e.g. Rieutord 2001). Then, being so different from equatorially concentrated
modes, they are not expected to follow the same regular spacings.

Another interesting issue is the difference between the modes of different equatorial symmetry.
We have seen that although the structure of the symmetric and anti-symmetric spectra is similar,
the frequency spectrum of the symmetric modes as a whole seems to evolve independently from the
anti-symmetric spectrum. The equatorial symmetry also influences the ``strength" of the avoided
crossings measured by the frequency separation at the closest frequency approach. As illustrated in Fig.~\ref{tout} (right panels),
avoided crossings between symmetric modes are always stronger than avoided crossings between anti-symmetric modes
since they remain further apart.

Modes undergoing an avoided crossing are particular because they have close frequencies and
similar eigenfunctions. As a consequence, both can be excited to observable levels by 
some excitation mechanism. They are therefore good candidates to explain the occurrence
of close frequencies in observed spectrum \citep{Br06} as well as the associated amplitude variations
induced by beating between the two close frequencies.

The most striking effect of the centrifugal force on the eigenfunction is the equatorial
concentration of the mode amplitude. Again, the study of the ray dynamics should help specify
the conditions in which the sound waves stay focused in the equatorial region.
As compared to the non-rotating case, the equatorial concentration
strongly modifies the integrated light visibility and in particular its variation
with respect to the mode degree and the inclination angle.
Interestingly enough, our results showing a global increase of the disk-integration factor as
the star is seen equator-on are compatible with observations of $\delta$ Scuti pulsations
which also suggest an increase of the pulsation amplitudes with $i$ \citep{Su02}.
Another finding of practical interest is that, for rapidly rotating stars,  the cancellation effect of the disk averaging
does no longer sharply decrease with the degree of the mode and also varies with the order of the mode.
Realistic calculations of the mode visibility including non-adiabatic calculations of the oscillation modes, stellar atmosphere models
as well as the gravity and limb darkening effect
will however be needed to draw firm observational conclusions.

Finally it should be reminded that the omission of the Coriolis force did not allow
a complete treatment of the rotational effects.
However, the effect of the 
Coriolis force vanishes for sufficiently large frequency (as the time scale
of the Coriolis acceleration $1/\Omega$ becomes much larger than the pulsation period)
while the modification of the equilibrium model
by centrifugal
force affects all frequencies. Therefore, the results presented here should be 
at least useful for the high frequency part of the acoustic spectrum in
rotating stars.
In a companion paper (Reese et al. 2006b), we extend the present results by taking into account
the Coriolis acceleration which, among other things,
allows us to specify the domain of validity of perturbative calculations.

\begin{acknowledgements}
We thank L. Valdettaro for his contribution on the numerical part of this work and B. Georgeot for fruitful discussions.
We also thank the referee for his constructive comments.
Numerical simulations have been performed with the computing resources of Institut du D\'eveloppement et des Ressources
en Informatique Scientifique (IDRIS, Orsay, France) and of CALcul en MIdi-Pyr\'en\'ees (CALMIP, Toulouse, France)
which are gratefully acknowledged.
\end{acknowledgements}

\appendix

\section{Linear operators in spheroidal coordinates}

Let us now express the linear operators involved in Eqs. \eq{oneII},
\eq{twoII}, \eq{threeII}, using the spheroidal coordinates given by Eq. \eq{eq:coor}. We need the general
expression of the divergence

\begin{equation}
\nab \cdot \vec{V} = \frac{1}{\sqrt{\mid g \mid}} {\partial}_i \lp \sqrt{\mid g \mid} V^i \rp
\end{equation}
\noi and the Laplacian:

\begin{equation}
\Delta \Phi = \frac{1}{\sqrt{\mid g \mid}} {\partial}_i \left( \sqrt{\mid g \mid} g^{in} {\partial}_n \Phi \right)
\end{equation}
where $\vec{V} = V^1 \vec{E}_1 + V^2 \vec{E}_2 + V^3 \vec{E}_3 = V_1
\vec{E}^1 + V_2 \vec{E}^2 + V_3 \vec{E}^3$ is written in the natural
basis $\vec{E}_i = \partial \vec{OM} / \partial x^i$ or the conjugated
basis $\vec{E}^i$ verifying $\vec{E}_i \cdot \vec{E}^j = \delta_{ij}$,
$g^{in}$ are the components of the metric tensor and $\mid \! g \! \mid$ is the
absolute value of metric tensor determinant.

\noi From these expressions, we derived the form of the following operators:
\begin{equation}
\vec{g}_0 \cdot \nab \equiv  (g^{1}_0 \vec{E}_1 + g^{2}_0 \vec{E}_2) \cdot {\partial}_{i} \; \vec{E}^i \equiv
g_0^1 {\partial}_{\zeta} + g_0^2 {\partial}_{\theta}
\end{equation}
\begin{equation}
r^2 \Delta \equiv  h_1 {\partial}_{\zeta \zeta}^2
- 2 h_2 {\partial}_{\theta \zeta}^2
+ h_4
{\partial}_{\zeta}
+ \Delta_{\theta \phi}
\end{equation}
\begin{equation}
r^2 {\cal L} \equiv - \lp h_1 \frac{g_0^2}{g_0^1} + 2 h_2 \rp
{\partial}_{\theta \zeta}^2
+ h_4 {\partial}_{\zeta}
- \lc h_1 {\partial}_{\zeta} \!\! \lp \frac{g_0^2}{g_0^1} \rp \rc {\partial}_{\theta}
+ \Delta_{\theta \phi}
\end{equation}
\begin{equation}
r^2 \nab \cdot (\; \bullet \; \vec{A_0}) \equiv (r^2 A_0^1)  {\partial}_{\zeta} +  (r^2 A_0^2) {\partial}_{\theta} + r^2 \nab \cdot \vec{A_0} [\mbox{id}]
\end{equation}
\noi where $\Delta_{\theta \phi}$ represent the horizontal part of the Laplacian in spherical
coordinates:

\begin{equation}
\Delta_{\theta \phi} \equiv {\partial}_{\theta \theta}^2 + \cot \theta {\partial}_{\theta} + \frac{1}{\sin^2\!\theta}
{\partial}_{\phi \phi}^{2},
\end{equation}
and

\begin{equation}
h_1 = \frac{r^2 + r_{\theta}^2}{r^2_{\zeta}}
\end{equation}
\begin{equation}
h_2 = \frac{r_{\theta}}{r_{\zeta}}
\end{equation}
\begin{equation}
h_3 = \frac{r}{r_{\zeta}}
\end{equation}
\begin{equation}
h_4 = \frac{1}{r_{\zeta}} \lc {\partial}_{\zeta} \!\! \lp \frac{r^2 + r_{\theta}^2}{r_{\zeta}} \rp
- \frac{1}{\sin \theta} {\partial}_{\theta}(r_{\theta} \sin \theta) \rc
\end{equation}
We recall that:
\begin{equation}
r = (1-\epsilon) \zeta + A(\zeta) \lp S(\theta) - 1+\epsilon \rp
\end{equation}
\noi where $S(\theta)$ describes the stellar surface.

\section{Coupling matrix}

The components of the sub-matrices which define the ODE system \eq{eq:sys} are specified below
using the functionals $\gi$ and $\gj$ defined in Eqs. \eq{eq:ii} and \eq{eq:jj}:

\begin{equation} \label{debut}
\mbox{A}_{33} \qquad \gi (h_1)
\end{equation}
\begin{equation}
\mbox{B}_{11} \qquad \gi (r^2 g_0^1)
\end{equation}
\begin{equation}
\mbox{B}_{21} \qquad \gi (h_4) - \gj \lp 2 h_2 + h_1 \frac{g_0^2}{g_0^1} \rp
\end{equation}
\begin{equation}
\mbox{B}_{22} \qquad \gi \lp h_1 \frac{c_0^2 N^2_0}{g_0^1} - r^2 A_0^1 \rp
\end{equation}
\begin{equation}
\mbox{B}_{33} \qquad \gi (h_4) -  \gj (2 h_2)
\end{equation}
\begin{equation}
\mbox{C}_{11} \qquad \gj (r^2 g_0^2)
\end{equation}
\begin{equation}
\mbox{C}_{12} \qquad -\gi (r^2 c_0^2 N^2_0)
\end{equation}
\begin{equation}
\mbox{C}_{21} \qquad - \llp \delta_{\lpp} - \gj \lc h_1 {\partial}_{\zeta} \!\! \lp \frac{g_0^2}{g_0^1} \rp \rc
\end{equation}
\begin{equation}
\mbox{C}_{22} \qquad \gi \lc h_1 {\partial}_{\zeta} \!\! \lp \frac{c_0^2 N^2_0}{g_0^1} \rp  - r^2 \nab \cdot \vec{A}_0 \rc - \gj (r^2 A_0^2)
\end{equation}
\begin{equation}
\mbox{C}_{31} \qquad - \gi (r^2 d_0)
\end{equation}
\begin{equation}
\mbox{C}_{32} \qquad \gi \lc r^2 (d_0 c_0^2 - \rho_0) \rc
\end{equation}
\begin{equation}
\mbox{C}_{33} \qquad - \llp \delta_{\lpp} + \gi (r^2 d_0)
\end{equation}
\begin{equation}
\mbox{D}_{21} \qquad - \gi \lp \frac{h_1}{g_0^1} \rp
\end{equation}
\begin{equation}
\mbox{D}_{22} \qquad \gi \lp \frac{h_1 c_0^2}{g_0^1} \rp
\end{equation}
\begin{equation}
\mbox{E}_{11} \qquad \gi (r^2)
\end{equation}
\begin{equation}
\mbox{E}_{12} \qquad -\gi (r^2 c_0^2)
\end{equation}
\begin{equation}
\mbox{E}_{21} \qquad -\gi \lc h_1 {\partial}_{\zeta} \!\! \lp \frac{1}{g_0^1} \rp \rc
\end{equation}
\begin{equation} \label{fin}
\mbox{E}_{22} \qquad \gi \lc r^2 + h_1 {\partial}_{\zeta} \!\! \lp \frac{c_0^2}{g_0^1} \rp \rc
\end{equation}
\noi where $\ell = m+ 2k$, $\ell' = m+ 2k'$ when applied to the
$\vec{\Xi}_{m}^{+}$ vector and $\ell = m+ 2k+1$, $\ell' = m+ 2k'+1 $
for $\vec{\Xi}_{m}^{-}$.

For polytropic model of index $N$, the quantities
describing the equilibrium can be expressed in terms of the dimensionless enthalpy
$H$ as follows:

\begin{equation} \label{eq:mod}
\begin{array}{ll}
\vec{g}_0 = \nab H & \vec{A}_0 = \lp 1 - \frac{N{\Gamma}_{1,0}}{N+1} \rp \nab H \\\\
c_0^2 = \frac{{\Gamma}_{1,0}}{N+1} H & c_0^2 N_0^2 = \lp 1 - \frac{N{\Gamma}_{1,0}}{N+1} \rp {\|\nab H\|}^2 \\\\
\rho_0 = \Lambda^N  H^N & d_0 = N \Lambda^N  H^{N-1},
\end{array}
\end{equation}
\noi where $\Lambda$ is such that
\begin{equation}
\Lambda = \frac{4 \pi G \rho_c R_e^2}{h_c}
\end{equation}
\noi where $h_c$ and $\rho_c$ are the dimensional enthalpy and density at the center of the polytropic model.

The components of the ODE, given by Eqs. \eq{debut} to \eq{fin}, can then
be expressed in terms of the
enthalpy and its derivatives, $H_{\zeta}, H_{\theta}, H_{\zeta \theta}, H_{\zeta \zeta}
H_{\theta \theta}$. This has been done in order to minimize the numerical error
in the calculation of these components.
The most useful expressions are:
\begin{equation}
g_0^1 = \frac{h_1 H_{\zeta} - h_2 H_{\theta}}{r^2}
\end{equation}
\begin{equation}
g_0^2 = \frac{- h_2 H_{\zeta} + H_{\theta}}{r^2}
\end{equation}
\begin{equation}
{\|\nab H\|}^2 = \frac{h_1 H_{\zeta}^2 -2 h_2 H_{\zeta} H_{\theta} + H_{\theta}^2}{r^2}
\end{equation}
\begin{equation}
h_1 \frac{c_0^2 N_0^2}{g_0^1} - r^2 A_0^1 = \lp 1 - \frac{N{\Gamma}_{1,0}}{N+1} \rp
\frac{H_{\theta}^2}{r_{\zeta}^2 g_0^1}
\end{equation}
\beqan
\partial_{\zeta} \!\! \lp \frac{g_0^2}{g_0^1} \rp &=& \frac{1}{\lp r^2 g_0^1 \rp^2} \lc
h_3^2 \lp H_{\zeta \theta} H_{\zeta}
- H_{\zeta \zeta} H_{\theta} \rp + \lp h_2 \partial_{\zeta} h_1 \right. \right. \nonumber \\
&& \left. \left. - h_1 \partial_{\zeta} h_2 \rp H_{\zeta}^2
- \partial_{\zeta} h_1  H_{\zeta} H_{\theta} + \partial_{\zeta} h_2 H_{\theta}^2 \rc
\eeqan{dale}
\begin{equation}
h_1 {\partial}_{\zeta} \!\! \lp \frac{c_0^2 N^2_0}{g_0^1} \rp  - r^2 \nab \cdot \vec{A}_0 =
- \lp 1 - \frac{N{\Gamma}_{1,0}}{N+1} \rp r^2 {\cal L}(H)
\end{equation}
\beqan
r^2 {\cal L}(H) &=& -h_1 \lc \frac{g_0^2}{g_0^1} H_{\theta \zeta} + \partial_{\zeta} \!\! \lp \frac{g_0^2}{g_0^1} \rp H_{\theta} \rc
-2 h_2 H_{\theta \zeta} \nonumber \\
&& + h_4 H_{\zeta} + \Delta_{\theta \phi} H
\eeqan{dole}
\begin{equation}
d_0 c_0^2 - \rho_0 = - \lp 1 - \frac{N{\Gamma}_{1,0}}{N+1} \rp \Lambda^N  H^N
\end{equation}
\beqan
\partial_{\zeta} \!\! \lp \frac{1}{g_0^1} \rp &=& - \frac{r^2}{\lp r^2 g_0^1 \rp^2} \lc h_1 H_{\zeta \zeta} -h_2 H_{\zeta \theta}
+ \lp \partial_{\zeta} h_1  \right.
\right.
\nonumber \\
&& \left. \left. - 2 h_1 / h_3 \rp H_{\zeta}  + \lp 2 h_2 / h_3 - \partial_{\zeta} h_2 \rp H_{\theta} \rc
\eeqan{dile}
\beqan
\partial_{\zeta} \!\! \lp \frac{c_0^2}{g_0^1} \rp = \frac{{\Gamma}_{1,0}}{N+1}
\lc H \partial_{\zeta} \!\! \lp \frac{1}{g_0^1} \rp + \frac{H_{\zeta}}{g_0^1} \rc
\eeqan{dule}

\noi where
\begin{equation}
\partial_{\zeta} h_1 = 2 \lp \frac{r_{\theta \zeta}}{r_{\zeta}} h_2 - \frac{r_{ \zeta \zeta}}{r_{\zeta}} h_1 + h_3 \rp
\end{equation}
\begin{equation}
\partial_{\zeta} h_2 = \frac{r_{\theta \zeta}}{r_{\zeta}} - \frac{r_{ \zeta \zeta}}{r_{\zeta}} h_2
\end{equation}

\section{Calculation of the disk-integration factor}

According to the definition of the disk-integration factor, Eq. \eq{eq:disk}, we are led to calculate integrals of the following form:
\beqan
I &=& \int\!\!\!\int_{S_v} F(\theta,\phi) {\bf d S} \cdot {\bf e_i} \\
 &=& \int\!\!\!\int_{S_v} G(\theta,\phi,i) d\mu d\phi
\eeqan{aaaa}
\noi where $\mu = \cos(\theta)$ and  $F(\theta,\phi)$ is the surface distribution of the eigenfunction obtained in the coordinate system 
\eq{eq:coor} in which the polar axis
is the rotation axis.
The integral is most simply calculated in the
coordinate system in which the polar axis is aligned with the direction of the observer.
This coordinate system results from a rotation of angle $i$ around the $y$ axis of the
original coordinate system, the
new angular variables being denoted $\theta'$ and $\phi'$.
To express $G$ in these coordinates, we use the formula relating the spherical harmonics in both systems:
\begin{equation}
\YL(\theta,\phi) =
\sum_{m' =-\ell}^{+\ell} d_{m m'}^{\ell}(i) \YLM(\theta ',\phi ')
\end{equation}
\noi where $d_{m m'}^{\ell}(i)$ do not generally have a simple form \citep{Ed60}.
Then, using the spherical harmonic expansion of $G$, we obtain:
\beqan
G &=& \sum_{\ell=0}^{+\infty}\sum_{m=-\ell}^{+\ell} G^\ell_m(i) \YL(\theta,\phi) \\
 &=& \sum_{\ell=0}^{+\infty}\sum_{m=-\ell}^{+\ell} \sum_{m'=-\ell}^{+\ell} G^\ell_m(i) d_{m m'}^{\ell}(i) \YLM(\theta',\phi')
\eeqan{aaab}
Then, integrating over the longitude $\phi'$, from $0$ to $2 \pi$, the terms
involving $\YLM(\theta',\phi')$ vanish if $m' \neq 0$. It follows that
\begin{equation}
I = 2 \pi \sum_{\ell=0}^{+\infty}\sum_{m=-\ell}^{+\ell} J_{\ell} G^\ell_m(i) \YTL(i)
\end{equation}
\noi where we used the following relations,
\begin{equation}
d_{m 0}^{\ell}(i) = \sqrt{\frac{4 \pi}{2 \ell + 1}} \YTL(i)
\end{equation}
\begin{equation}
d\mu d\phi =  d\mu' d\phi' \;\; \mbox{where $\mu' = \cos \theta'$}
\end{equation}
\noi and defined $J_{\ell}$ as,
\beqan
J_{\ell} &=& \sqrt{\frac{4 \pi}{2\ell+1}} \int_0^{1} \hat{Y}^0_\ell(\mu') d\mu' \\
&=& \left\{ \begin{array}{lll}
         0 & \mbox{if $\ell$ is even and $\ell \neq 0$},\\
         1 & \mbox{if $\ell=0$},\\
	 (-1)^{\frac{\ell-1}{2}} \frac{1.3...(\ell-2)}{2.4...(\ell+1)} & \mbox{if $\ell$ is odd and $\ell \neq 1$}, \\
	 \frac{1}{2} & \mbox{if $\ell=1$}.\end{array} \right.
\eeqan{aaac}
Because of axial axial symmetry, the function to integrate reads 
\begin{equation}
F(\theta,\phi) = W(\theta) e^{im\phi}.
\end{equation}
Then, from the expression of the vector $\vec{dS}$ at the star's surface:
\begin{equation}
\vec{dS} = \partial_{\theta} \vec{OM} \times \partial_{\phi} \vec{OM} d\theta d\phi = \vec{E}_2 \times \vec{E}_3 d\theta d\phi = \sqrt{g} \vec{E}^1 
d\theta d\phi
\end{equation}
we deduce that
\begin{equation}
G = r A(\theta,\phi,i) W(\theta) e^{im\phi} \\
\end{equation}
\noi where
\beqan
A(\theta,\phi,i) &=& r r_{\zeta} \vec{E^1} \cdot \vec{e_i} = \sqrt{r^2 + r^2_{\theta}} \vec{e^s} \cdot \vec{e_i} \\
& = &r \lp \sin\theta\cos\phi\sin i + \cos\theta \cos i \rp +
\nonumber \\
&&
r_{\theta} \lp \sin\theta \cos i - \cos \theta \cos\phi \sin i \rp \\
& = & \cos i \frac{d}{d \theta} \lp r \sin\theta \rp -\sin i \cos \phi \frac{d}{d \theta} \lp r \cos \theta \rp
\eeqan{aaad}
\noi where $\vec{e^s}$ denotes the unit vector perpendicular to the surface, $r$ and$r_{\theta}$ are calculated at the
star surface $\zeta=1$. Thus the dependency of $G$ on $i$, $\phi$ and $\theta$ can be specified
as follows:
\beqan
G &=& A(\theta) \cos i e^{im\phi}  - B(\theta) \sin i \cos \phi e^{im\phi}
\eeqan{aaae}
\noi where 
\beqan
A &=& r \frac{d}{d \theta} \lp r \sin\theta \rp W(\theta) \\
B &=& r \frac{d}{d \theta} \lp r \cos\theta \rp W(\theta)
\eeqan{aaaf}
\noi It follows that
\begin{equation}
G^\ell_k = 0 \;\;  \mbox{if $k \neq m-1,m,m+1$}
\end{equation}
\noi so that the integral now reads:
\beqan
I/2\pi &=& I_{m-1}+  I_m + I_{m+1} \;\; \mbox{where} \\
I_m &=& \cos i \hat{A}_m(i) \\
I_{m-1} &=& - \frac{\sin i}{2} \hat{B}_{m-1}(i) \\
I_{m+1} &=& - \frac{\sin i}{2} \hat{B}_{m+1}(i) 
\eeqan{aaag}
where $\hat{A}_m$ denotes:
\beqan
\hat{A}_m(i) &=& \sum_{\ell=|m|}^{+\infty} J_\ell A^\ell_m \YTL(i) \\
A^\ell_m &=& 2\pi \int_0^{\pi} A(\theta) \YTL(\theta) \sin \theta d \theta
\eeqan{aaah}
\noi the $\hat{B}_m$ terms being defined accordingly.

Note that for modes which are equatorially anti-symmetric and axisymmetric ($m=0$),
$\hat{A}_0(i)=J_0 A^0_0 \hat{Y}^0_0(i)$ and $\hat{A}_1(i) = \hat{A}_{-1}(i) = 0$, thus
the integral $I$ reduces to:
\beqan
I = 4 \pi \sqrt{\pi} A^0_0 \cos(i).
\eeqan{aaai}
\end{document}